\documentclass[a4paper,english,reprint, pra, letterpaper]{revtex4-1}
\usepackage[T1]{fontenc}
\usepackage[utf8]{inputenc}
\usepackage{amsmath}
\usepackage{amssymb}
\usepackage{graphicx}

\makeatletter

\pdfpageheight\paperheight
\pdfpagewidth\paperwidth

\usepackage{lmodern}
\usepackage[T1]{fontenc}

\makeatother

\usepackage{babel}
\begin{document}

\title{Detuning-dependent Properties and Dispersion-induced Instabilities
of Temporal Dissipative Kerr Solitons in Optical Microresonators}

\author{Erwan Lucas}

\affiliation{École Polytechnique Fédérale de Lausanne (EPFL) -- Institute of
Physics, Lausanne, CH-1015, Switzerland.}

\author{Hairun Guo}

\affiliation{École Polytechnique Fédérale de Lausanne (EPFL) -- Institute of
Physics, Lausanne, CH-1015, Switzerland.}

\author{John D. Jost}

\affiliation{École Polytechnique Fédérale de Lausanne (EPFL) -- Institute of
Physics, Lausanne, CH-1015, Switzerland.}

\author{Maxim Karpov}

\affiliation{École Polytechnique Fédérale de Lausanne (EPFL) -- Institute of
Physics, Lausanne, CH-1015, Switzerland.}

\author{Tobias J. Kippenberg}
\altaffiliation{Corresponding author: tobias.kippenberg@epfl.ch}

\affiliation{École Polytechnique Fédérale de Lausanne (EPFL) -- Institute of
Physics, Lausanne, CH-1015, Switzerland.}

\date{\today}
\begin{abstract}
Temporal-dissipative Kerr solitons are self-localized light pulses
sustained in driven nonlinear optical resonators. Their realization
in microresonators has enabled compact sources of coherent optical
frequency combs as well as the study of dissipative solitons. A key
parameter of their dynamics is the effective-detuning of the pump
laser to the thermally- and Kerr-shifted cavity resonance. Together
with the free spectral range and dispersion, it governs the soliton-pulse
duration, as predicted by an approximate analytical solution of the
Lugiato-Lefever equation. Yet, a precise experimental verification
of this relation was lacking so far. Here, by measuring and controlling
the effective-detuning, we establish a new way of stabilizing solitons
in microresonators and demonstrate that the measured relation linking
soliton width and detuning deviates by less than 1\% from the approximate
expression, validating its excellent predictive power. Furthermore,
a detuning-dependent enhancement of specific comb lines is revealed,
due to linear couplings between mode-families. They cause deviations
from the predicted comb power evolution, and induce a detuning-dependent
soliton recoil that modifies the pulse repetition-rate, explaining
its unexpected dependence on laser-detuning. Finally, we observe that
detuning-dependent mode-crossings can destabilize the soliton, leading
to an unpredicted soliton breathing regime (oscillations of the pulse)
that occurs in a normally-stable regime. Our results test the approximate
analytical solutions with an unprecedented degree of accuracy and
provide new insights into dissipative-soliton dynamics. 
\end{abstract}


\maketitle
\global\long\def\mgf{\mathrm{{MgF_{2}}}}
\global\long\def\sech{\operatorname{sech}}

\section{Introduction}

 Dissipative Kerr-cavity Solitons (DKS) are self-localized pulses
of light that can be excited in coherently-driven nonlinear optical
resonators. Following earlier studies of externally-induced dissipative
solitons in fiber cavities \cite{Leo2010}, they were shown to spontaneously
form in microresonators \cite{Herr2013}. From an applied perspective,
DKS generation in microresonators enables high-repetition rate sources
of ultrashort pulses, producing coherent, broadband optical ``Kerr''
frequency combs \cite{Kippenberg2011}. Kerr frequency combs are generated
by coupling a strong continuous wave laser into a nonlinear microresonator
that converts the initial frequency into a set of equidistant comb
lines via a cascade of parametric effects \cite{Del'Haye2007}. With
proper tuning of the pump laser, these processes result in the formation
of DKS in the cavity sustained via the double balance between cavity
loss and parametric gain, as well as dispersion and Kerr nonlinearity
\cite{Leo2010,Coen2013,Herr2013,Matsko2011b,Akhmediev2003}. These
DKS-based frequency combs have been demonstrated in several microresonator
platforms, enabling on-chip photonic integration \cite{Herr2013,Brasch2015,Yi2015,Joshi2016}.
Compared to other optical frequency comb platforms, DKS combs extend
the repetition rate to the microwave and milli-meter wave domain,
while simultaneously providing wide bandwidth and compact form factor.
They have already been successfully used for coherent terabit communications
\cite{Marin-Palomo2016}, microwave-to-optical phase coherent links
\cite{Jost15,DelHaye2016,Brasch2017}, and the generation of low noise
microwaves \cite{Liang2015a}. The interplay of the fundamental aspects
of soliton physics and their applications has shown the suitability
of the microresonator platform to study soliton properties. A recent
demonstration evidenced how soliton Cherenkov radiation in a dispersion-managed
resonator \cite{Jang2014,Akhmediev1995a,Brasch2015} can extend the
frequency comb bandwidth, enabling self-referencing without external
broadening \cite{Brasch2017}.

Fundamentally, the dynamics of the DKS rely on the resonator properties
and two external parameters of the pump laser: the power and the detuning
to the pumped resonance. An analytical estimate \cite{Herr2013,Coen2013a}
predicts that the soliton duration (and thus the comb bandwidth) only
depends on the resonator free spectral range, dispersion, and detuning.
While the former two parameters are readily accessible and measurable
with high precision, the detuning of the\textit{\emph{ nonlinear syste}}m
is more challenging to determine, in particular since microresonators
are susceptible to thermal nonlinearities \cite{Carmon2004,DelHaye2015a}.
Here, we apply a recently introduced method \cite{Matsko2015,Guo2016}
enabling detuning measurement, to carry out a controlled study of
the effect of the detuning on the properties of a single soliton in
a crystalline magnesium fluoride ($\mathrm{MgF_{2}}$) resonator,
and perform a careful comparison of the measurements to the theoretical
predictions. This is achieved via a feedback-stabilization of the
detuning parameter, which ensures the stability over the measurement
duration and and enabled long-term soliton stabilization. The results
show very good agreement between the soliton pulse bandwidth and the
analytical approximation that deviate by less than 1\%. Local features
in the resonator dispersion caused by coupling of other spatial mode
families induce detuning-dependent spectral features, which are shown
to cause a soliton recoil, and affect the repetition rate as well
as the total comb power. Unexpectedly, mode crossings are further
shown to alter the soliton stability, leading to a ``breathing''
regime in which the soliton amplitude and width oscillate. This soliton
breathing occurs at a detuning range, where the solitons are expected
to be stable. Beyond elucidating the detuning dependence of temporal
solitons, this work, to the best of the authors knowledge, constitutes
a direct experimental verification of the DKS models with an accuracy
that has not been attained in previous studies of this class of solitons.

\section{Analytical description}

The complex dynamics of a continuous-wave (CW) laser-driven nonlinear
optical microresonator can be described both in the frequency and
time domains, via coupled mode equations \cite{Herr2012} or via a
spatiotemporal description \cite{Chembo2013,Wabnitz1993}. In the
time domain, the equation of motion for the envelope of the cavity
field is given by:
\begin{align}
\frac{\partial A}{\partial t} & =-\left(\frac{\kappa}{2}+i\delta\omega\right)A+i\dfrac{D_{2}}{2}\frac{\partial^{2}A}{\partial\phi^{2}}+ig_{0}|A|^{2}A\nonumber \\
 & \quad+\sqrt{\frac{\eta\kappa P_{in}}{\hbar\omega_{0}}}
\end{align}
where $\kappa$ denotes the loaded resonator linewidth ($Q=\omega_{0}/\kappa$,
loaded quality factor), $\eta=\kappa_{ex}/\kappa$ the coupling coefficient,
$P_{in}$ the pump power, $\omega_{0}$ the pumped resonance frequency
(thermally shifted) and $\delta\omega=\omega_{0}-\omega_{p}$ is the
detuning of the pump laser to this resonance. The dispersion of the
resonator is described by expressing the resonance frequency as a
function of the mode number $\mu$ (relative to the pumped mode) as
$\omega_{\mu}=\omega_{0}+\mu D_{1}+\mu^{2}D_{2}/2$, where $D_{1}$
correspond to the FSR in rad/s and $D_{2}$ relates to the GVD parameter
$\beta_{2}$ ($D_{2}=-\beta_{2}D_{1}^{2}c/n_{0}$). The nonlinearity
is described via the (per photon Kerr frequency shift) coefficient
$g_{0}=\hbar\omega_{0}^{2}cn_{2}/n_{0}^{2}V_{\mathrm{eff}}$, with
the refractive index of $\mathrm{MgF_{2}}$ $n_{0}$, nonlinear refractive
index $n_{2}$, and the effective cavity nonlinear volume $V\mathrm{_{eff}}=A_{\mathrm{eff}}L$
($A_{\mathrm{eff}}$ is the effective nonlinear optical mode area
and $L$ the circumference of the cavity). Under suitable normalization,
the above equation has been shown to be equivalent to the Lugiato-Lefever
equation (LLE) that originally described spatial pattern formation
in diffractive cavities \cite{Herr2013,Chembo2013,Matsko2011b,Lugiato1987}.
For anomalous group velocity dispersion ($D_{2}>0$), there exist
stable solutions consisting of DKS on top of a weak continuous field.
The approximate expression for the soliton component yields a hyperbolic
secant pulse such that for a single soliton in the microresonator,
the comb power spectral envelope follows a $\sech^{2}$ spectral profile
\cite{Herr2013,Coen2013a}:
\begin{align}
P(\mu) & \approx\frac{\pi}{2}\frac{\eta}{Q}\frac{D_{2}}{D_{1}}\frac{n_{0}A_{\mathrm{eff}}}{n_{2}}\,\sech^{2}\left(\frac{\pi\tau}{2}\mu\omega_{r}\right),\label{eq:SolSpectralPow}\\
\tau & \approx\frac{1}{D_{1}}\sqrt{\dfrac{D_{2}}{2\delta\omega}},\label{eq:SolDuration}
\end{align}
where $\omega_{r}$ is the comb repetition rate and $\tau$ the pulse
duration (corresponding pulse FWHM $\tau_{\mathrm{FWHM}}=2\mathrm{\,acosh}(\sqrt{2})\,\tau$).
Therefore, following this approximation, the soliton pulse duration
is only determined by \textit{three frequencies}, while the other
cavity properties determine the soliton's power levels.

 Eq. \eqref{eq:SolDuration} is at the core of several recent works
on soliton, as in \cite{Yi2015}, where it was employed to replace
$\delta\omega$ dependencies with $\tau$. A direct verification of
this approximation with experiment would further consolidate the validity
of such approach. Surprisingly, although simulations of the LLE and
comparisons to soliton experiments have rapidly advanced in recent
years, the fundamental test of \eqref{eq:SolDuration} in microresonators
has not been directly performed, due to lack of direct access to $\delta\omega$
in the driven nonlinear system in the presence of solitons. In microresonators,
photo-thermal and Kerr effects play a key role \cite{Carmon2004}.
When tuning the laser across a resonance to obtain a soliton state,
the thermal effect shifts the cavity resonance from its original cold
position \cite{Herr2013,Brasch2015}, making it difficult to precisely
infer the \textit{effective} laser detuning from this `hot' cavity
resonance.

In addition, this detuning not only determines the soliton duration,
but also if the soliton can be sustained. The soliton is indeed supported
in the cavity over a limited range of effective red-detuning ($0<\delta\omega<\delta\omega_{\mathrm{max}}$)
referred to as soliton existence range. Therefore, thermal drifts
of the microresonator cavity can cause the effective detuning to walk
outside of these limits, leading to the decay of the soliton state.

\section{Results}

\begin{figure}
\includegraphics[width=1\columnwidth]{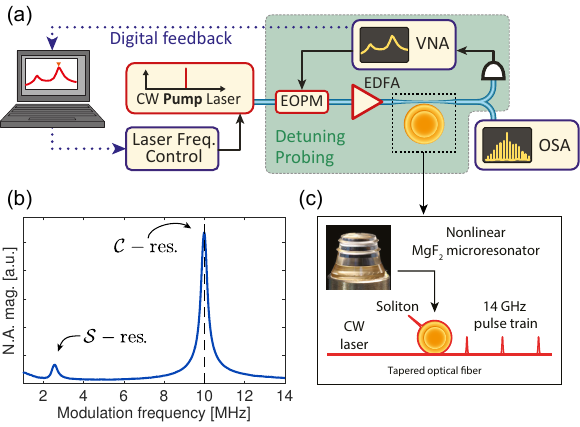}

\caption{\textbf{\label{fig:Setup_stabilization}Kerr comb generation, probing
and stabilization. (a)} Experimental setup: Vector Network Analyzer
(VNA), Electro-Optic Phase Modulator (EOPM), Erbium Doped Fiber Amplifier
(EDFA), Optical Spectrum Analyzer (OSA). \textbf{(b)} Double-resonance
cavity transfer function in the soliton state, as measured on the
VNA. The frequency of the $\mathrm{\mathcal{C}}$-resonance indicates
the pump-resonator detuning. \textbf{(c)} Principle of microresonator
frequency comb generation and formation of dissipative Kerr solitons.}
\end{figure}
\begin{figure}
\includegraphics[width=1\columnwidth]{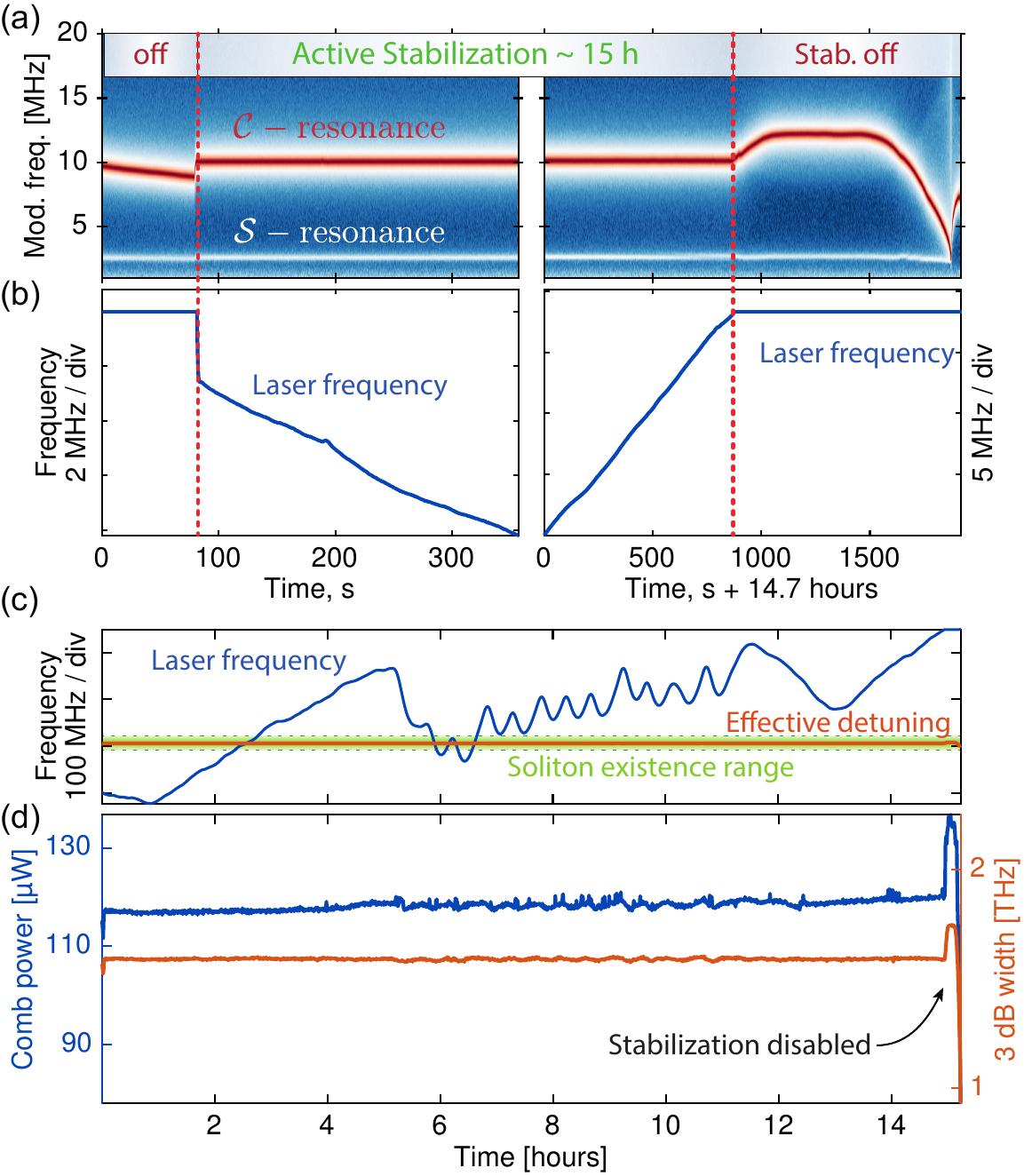}

\caption{\textbf{\label{fig:Stabilization_results}Effective detuning stabilization
of a dissipative Kerr soliton state. (a-b) }Close-in view of the lock
enabling and disabling. The colormaps in \textbf{(a)} show the concatenated
set of acquired VNA traces used to determine the detuning. The plots
in \textbf{(b) }trace the pump frequency. If the lock is enabled,
the laser is tuned to keep the effective detuning at a fixed value.
When the lock is disabled, the laser frequency is fixed, but the soliton
is lost after 17 min.\textbf{ (c-d)} Stabilization and continuous
soliton measurement over 15 h. \textbf{(c)} The blue line indicates
the evolution of the pump laser frequency when tracking the microresonator
resonance, which is measured by counting the heterodyne beat of the
pump with an ultra-stable laser. The temperature drifts of the microresonator
cavity are the main source of variations and the slow oscillations
are caused by the air conditioning. The red line indicates the stabilized
effective detuning (at 10~MHz) that remains within the soliton existence
range. \textbf{(d)} The comb power and the 3 dB bandwidth (obtained
by fitting the optical spectra) are stabilized when the laser compensates
the drifts.}
\end{figure}
\begin{figure*}
\includegraphics[width=1\textwidth]{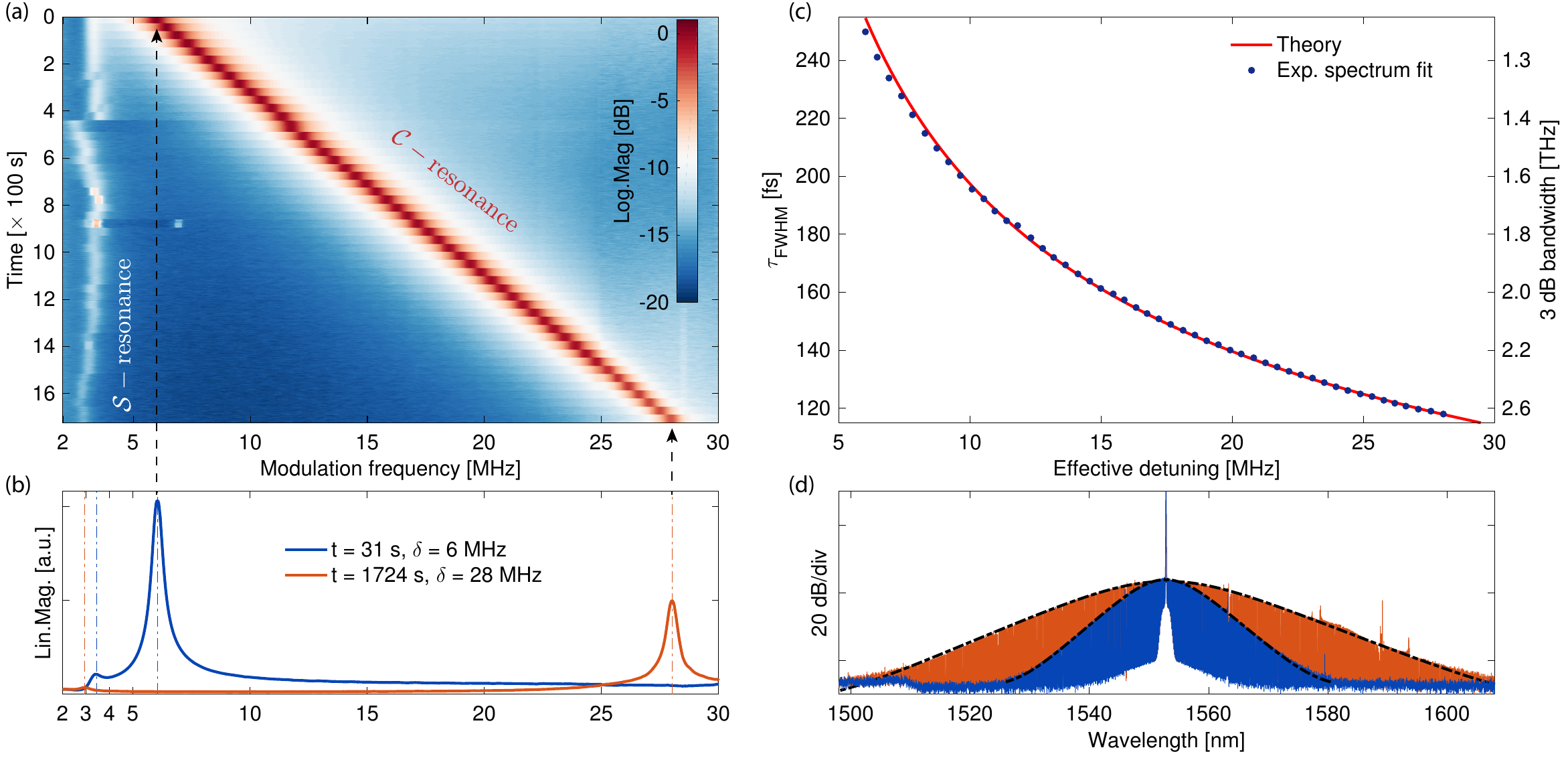}

\caption{\textbf{\label{fig:DeffSweepMap}Tuning of the effective detuning
and evolution of the soliton duration. (a)} Map showing the evolution
of the modulation response (log scale) as the effective detuning is
swept. The detuning is stabilized at each step. \textbf{(b) }The observed
VNA traces at the extrema of the effective detuning ($\delta\omega$).\textbf{
(c)} The measured soliton full width at half maximum (derived from
a $\protect\sech^{2}$ fit) is plotted versus the detuning (blue dots)
with comparison to the expression in Eq. \eqref{eq:SolDuration} (red
line). \textbf{(d)} Corresponding spectra at the limits of the sweep.
As expected, the comb bandwidth increases with larger effective detuning.
The black lines mark a $\protect\sech^{2}$ fit of the combs. \textbf{\label{fig:CombBW_detun}} }
\end{figure*}
\begin{figure}
\includegraphics[width=1\columnwidth]{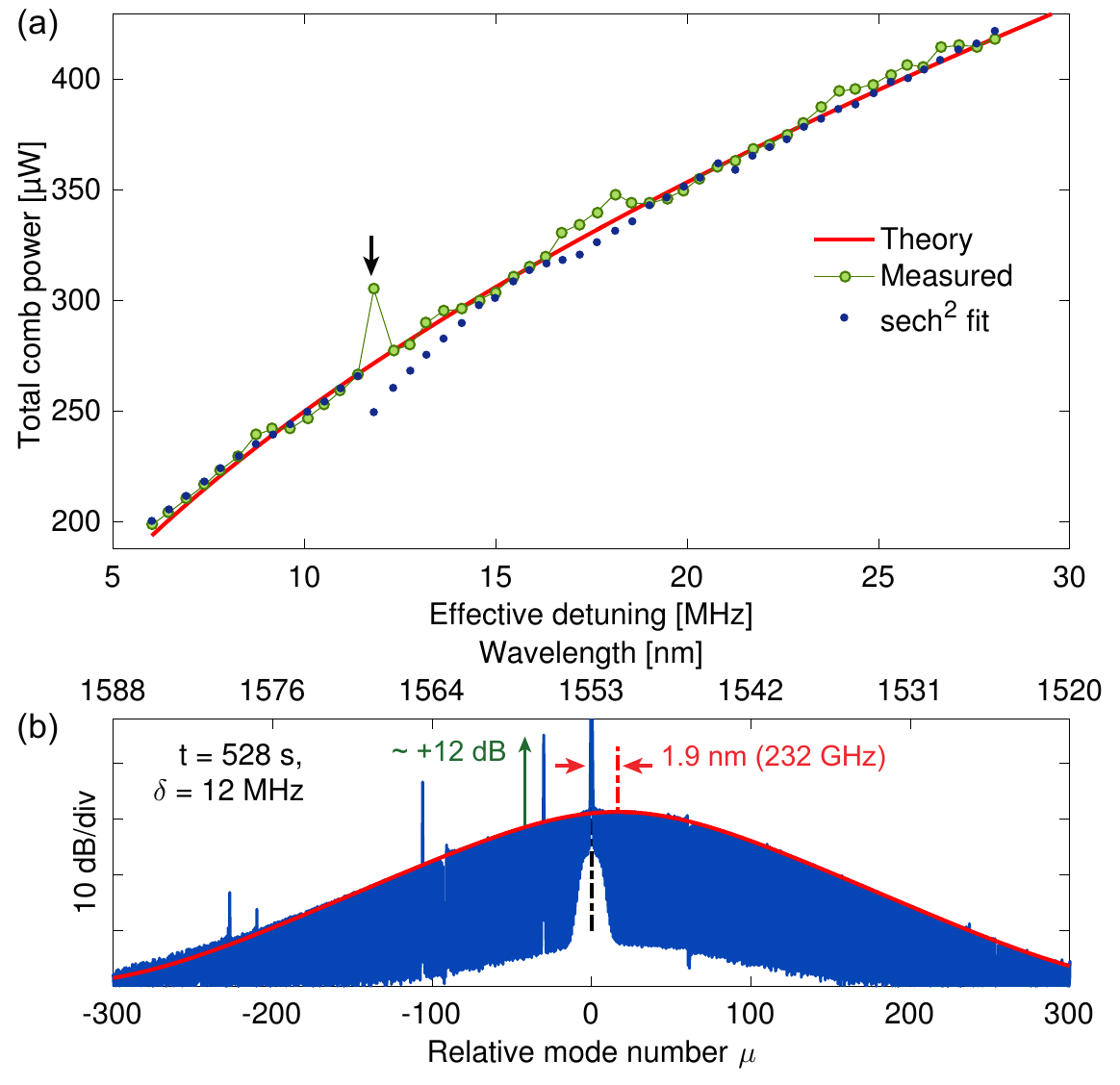}

\caption{\textbf{Evolution of the soliton power. (a)} Evolution of the measured
comb power with the effective detuning (green dots), compared to Eq.
\eqref{eq:SolAvgPow}, and the estimated power in the soliton component
(blue dots, derived from the $\protect\sech^{2}$ fit). \textbf{(b)}
Comb spectrum corresponding to the arrow in \textbf{a}. The black
dashed line marks the pump position ($\mu=0$). Two strong avoided
mode crossing are visible at $\mu=-31$ and $\mu=-106$, and induce
a shift of the $\protect\sech^{2}$ centroid from the pump toward
shorter wavelength, marked by the red arrows.\label{fig:CombPow_detun}}
\end{figure}
\begin{figure*}
\includegraphics[width=1\textwidth]{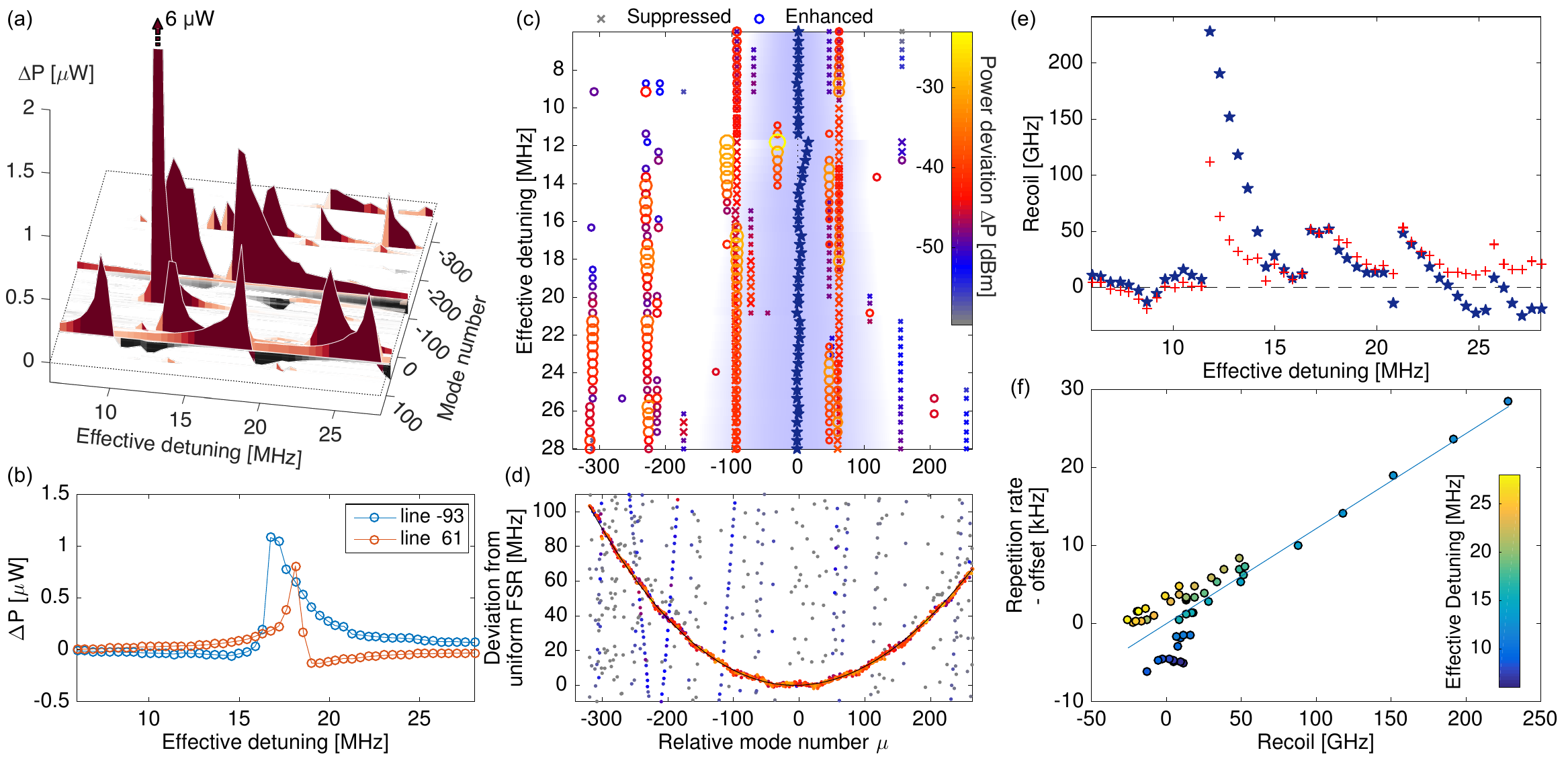}

\caption{\textbf{\label{fig:AvoidedXssing}Effect of detuning dependent avoided
mode crossings on the soliton frequency comb.} \textbf{(a)} Map of
$\Delta P$ indicating the spurs and dips in the spectrum after subtracting
the fitted $\protect\sech^{2}$ soliton envelope. \textbf{(b)} Section
of the $\Delta P$ map showing the evolution of the power deviation
for the comb line +61 and -93 (relative to the pump) \textbf{(c)}
Representation of the peaks in the $\Delta P$ map, in logarithmic
units, showing the evolution of the intensity spurs caused by avoided
mode crossings. The lines higher than the $\protect\sech^{2}$ envelope
(enhanced) are marked with a dot, the lines lower (suppressed) with
a cross. The blue stars mark the comb centroid $\Omega$. The shaded
blue region indicates the comb 3 dB width. When lines are strongly
enhanced, the comb centroid shift away from them.\textbf{ (d)} Measured
frequency dispersion of the mode family supporting the soliton. A
quadratic fit yields $D_{1}/2\pi=14.0938$~GHz and $D_{2}/2\pi=1.96$~kHz.
Multiple mode families with a different FSR exist in the resonator
and cross the family of interest, inducing small periodic disruptions
on the dispersion. \textbf{(e) }Evolution of the soliton recoil. The
blue stars result from the fit of the optical spectrum, while the
red crosses mark the estimated recoil using \eqref{eq:RecoilEstimate}.
\textbf{(f)} The repetition rate frequency is strongly correlated
with the recoil. This enables the determination of the dispersion
parameter as given by the slope ($D_{2}/D_{1}$). The offset on the
repetition rate is 14.094005~GHz.}
\end{figure*}
\begin{figure}
\includegraphics[width=0.8\columnwidth]{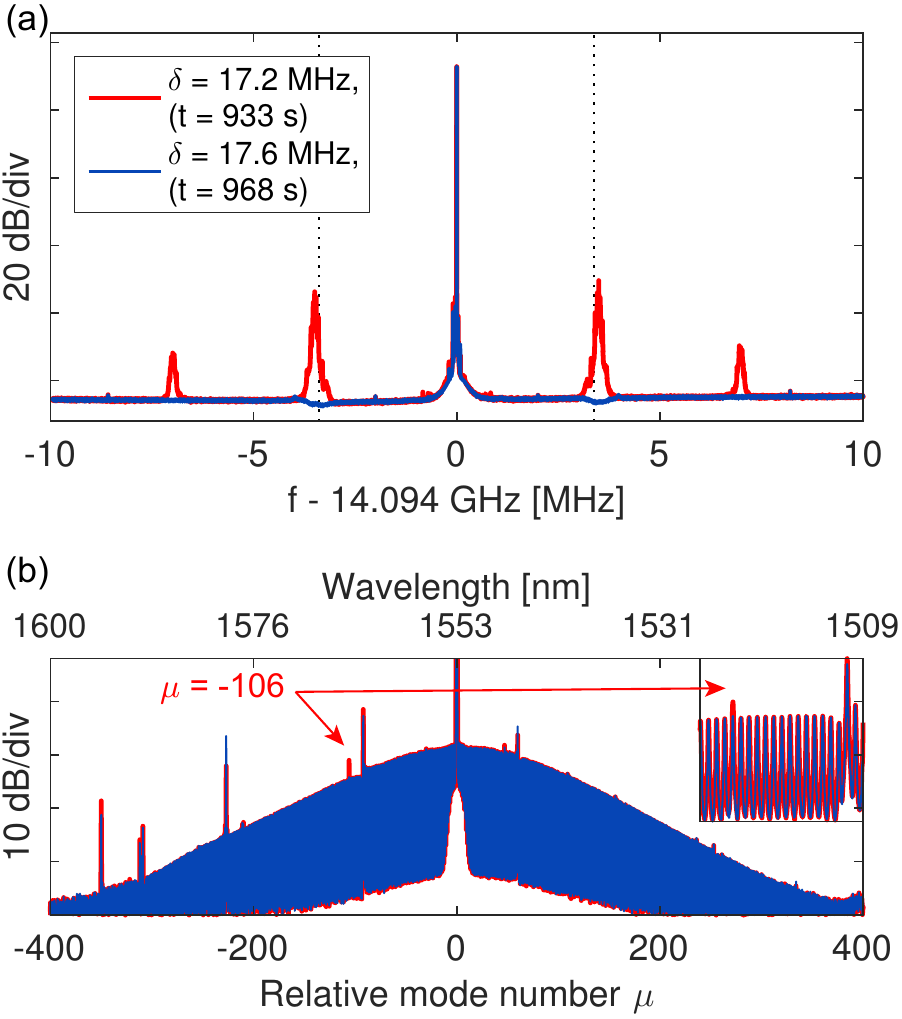}

\caption{\textbf{Avoided mode crossing induced soliton breathing }(a) RF spectrum
of the repetition rate for two adjacent detuning steps $\delta\omega/2\pi=17.2\ \text{{and}}\ 17.6$~MHz
(resolution bandwidth 1 kHz). In the first case, modulations sidebands
appear on the repetition beatnote, with a frequency of $\sim3.5$
MHz, closely matching the $\mathcal{S}$-resonance frequency measured
on the VNA (indicated by the dashed lines). This is typically indicative
of a soliton breathing. (b) Corresponding optical spectrum comparison.
The red (blue) trace corresponds to the soliton breathing (stable).
The breathing seems to correlate with the excitation of the mode $\mu=-106$.}
\end{figure}

\subsection{Effective detuning probing and stabilization of a dissipative Kerr
soliton state}

In order to study the soliton properties as a function of the effective
detuning, this parameter must be measured, stabilized, and tuned in
a controlled way. We recently demonstrated a way to probe the effective
detuning within the soliton state \cite{Guo2016}, akin to techniques
employed in ultrafast lasers \cite{Wahlstrand2007,Lee2014}. The underlying
idea is to frequency-sweep weak phase-modulation sidebands imprinted
onto the pump laser and record the resulting amplitude modulation
of the optical power coming out of the cavity. The sweep is generated
with a Vector Network Analyzer (VNA) and converted to a phase modulation
on the laser with an Electro-Optical Modulator (EOM). After the resonator,
the corresponding amplitude modulation is recorded on a photodiode
and demodulated by the VNA (see Fig. \ref{fig:Setup_stabilization}a).
When solitons propagate in the cavity, the system's transfer function
exhibits a \emph{double-resonance} feature, related to the strong
bistability of the cavity that supports both a weak CW background
and high intensity solitons. A first small peak at low modulation
frequencies is observed ($\mathrm{\mathcal{S}}$-resonance) that relates
to a resonance of the soliton and is weakly dependent on the detuning.
A second stronger peak ($\mathrm{\mathcal{C}}$-resonance) is also
measured, whose frequency corresponds to the effective detuning $\delta\omega$
of the pump laser to the optical resonance of the microresonator,
when $\delta\omega\gg\kappa$. The soliton existence range can be
determined easily with this probing technique by detuning the laser
until the soliton is lost. We measured it to range from $\delta\omega/2\pi\sim2$
MHz to $\sim30$ MHz, which corresponds to an effective laser-cavity
detuning of $\delta\omega/\kappa\sim160$ times the resonance linewidth.
This is enabled by the strong pumping of the resonator, that is $\sim140$
times above the parametric threshold ($P_{in}\approx215$~mW, intrinsic
linewidth $\kappa_{0}/2\pi\approx100$~kHz, $\eta\approx0.43$) \cite{Herr2012}.

We implemented a digital feedback-stabilization of the effective detuning,
as shown in Fig. \ref{fig:Setup_stabilization}a. The response of
the system is measured with the VNA (sweep time $\sim100$~ms) and
recorded with a computer. The detuning value is identified by detecting
the $\mathrm{\mathcal{C}}$-resonance frequency with a peak detection
algorithm, and the program determines the required feedback to apply
to the pump laser frequency to stabilize the detuning to a given value.
The overall feedback is slow ($\sim10$~Hz) but sufficient to compensate
the thermal drift, which is the main source of variations. This method
enabled the long term stabilization of a single soliton in the crystalline
microresonator over 15 h as presented on Figure \ref{fig:Stabilization_results}.
Over this period, the laser frequency was adjusted by more than 350~MHz,
which represents over ten times the existence range of the soliton.
The active compensation maintained the effective detuning fixed at
10~MHz and stabilized the comb bandwidth  (Fig. \ref{fig:Stabilization_results}c,e).
However, the parameters of the resulting frequency comb are not stabilized,
since the cavity FSR drifts thermally and so does the pulse repetition
rate. To highlight the effect of the stabilization, the lock was disabled
on purpose after $\sim15$ h and the thermal drifts caused the comb
properties to drift until the soliton state decayed after 17~min.

\subsection{Study of the detuning-dependent dissipative Kerr soliton duration}

In order to study the dependence of the soliton on the effective detuning,
this parameter was swept by changing the set-point in the computer.
Figure \ref{fig:DeffSweepMap}a shows a sweep of the effective detuning
from 6 to 28 MHz, in 50 steps. At each step, once the detuning was
stabilized, an optical spectrum was acquired (OSA scan time $\sim30$
s) and the comb average power (after suppressing the pump with a narrowband
fiber Bragg grating) was measured with a photodiode, before moving
to the next detuning value. At the same time, $\omega_{r}$ was measured
with a frequency counter after photo-detection and down-mixing. The
overall measurement duration is $\sim30$ min and the active detuning
stabilization is required to counteract the environmental drifts.
Each optical spectrum was fitted with the following expression:
\begin{equation}
A\,\mathrm{sech^{2}}\left(\dfrac{\mu\omega_{r}-\Omega}{B}\right),\label{eq:FitExpr}
\end{equation}
where $\mu$ is the relative mode number, $\omega_{r}$ the repetition
rate of the comb, $B=2/(\pi\:\tau)$ the bandwidth, $A$ the power
of the central comb line and $\Omega$ the spectral shift of the comb
centroid from the pump.

The presented method enables a precise comparison between the measured
comb properties and the theoretical predictions. The dispersion properties
of the resonator were measured experimentally via frequency comb assisted
scanning laser spectroscopy \cite{Herr2013b,Del'Haye2009} and shown
in Figure \ref{fig:AvoidedXssing}d. (the corresponding dispersion
parameters are $D_{1}/2\pi=14.094$~GHz, $D_{2}/2\pi=1.96$~kHz,
$D_{3}/2\pi=-1.39$~Hz). The soliton spectral bandwidth (and deduced
pulse duration) obtained experimentally is compared with the approximate
expression Eq. \eqref{eq:SolDuration}, using the measured dispersion
and detuning parameters (Fig. \ref{fig:CombBW_detun}a). We observe
an excellent agreement of the two curves (normalized RMS deviation
of 0.8 \%) supporting the validity of the approximation. The results
also show that the soliton duration can be tuned by more than a factor
of 2 by changing the detuning.

\subsection{Study of the detuning-dependent mode crossings and soliton recoil}

The relation between average power of the comb and detuning is obtained
by integrating Eq.~\eqref{eq:SolSpectralPow}:
\begin{equation}
\bar{P}=\dfrac{2\eta A_{\mathrm{eff}}n_{0}\kappa}{n_{2}\omega_{0}D_{1}}\sqrt{2D_{2}\delta\omega}.\label{eq:SolAvgPow}
\end{equation}
The evolution of the measured comb power, shown in Figure \ref{fig:CombPow_detun}a,
follows the trend of the previous equation, but significant discrepancies
are observed at some detuning values, such as for $\delta\omega/2\pi=12$
MHz, where a large spike in the comb power is measured. Integrating
the fit expression \eqref{eq:FitExpr} reveals that the power in the
soliton is reduced at these points (blue dots in Fig. \ref{fig:CombPow_detun}a).
The corresponding spectrum exhibits specific comb lines that are strongly
enhanced (Fig. \ref{fig:CombPow_detun}b). This effect is typically
caused by avoided mode crossing, where the coupling between two spatial
mode families causes a local disruption in the resonator dispersion,
leading to a modification of the phase matching condition between
the pump to a sideband mode. This is associated with an enhancement
or suppression of the comb generation at the crossing position \cite{Herr2013b,Grudinin2013b,Zhou2015}.
The excess power in certain lines (spikes) makes the frequency comb
asymmetric, which induces a recoil – i.e. a shift in the soliton center
frequency with respect to the pump – in the opposite direction, in
order to keep the spectral center of mass invariant \cite{Akhmediev1995a,Milian2014,Matsko2016a}.
In the time domain, the spike beats with the pump laser, leading to
an oscillating intracavity background. The soliton(s) are then trapped
on this oscillating pattern, creating a bound state \cite{Wang2016}.

The evolution of the mode crossing features with the laser detuning
is further investigated in Figure \ref{fig:AvoidedXssing}. Interestingly,
the measured dispersion of the mode family supporting the soliton
does not exhibit strong disruptions (see Fig. \ref{fig:AvoidedXssing}d),
instead we observe periodic crossings with a mode family having a
different FSR . We detect the mode crossings features in the comb
spectrum by first subtracting the $\sech^{2}$ fit, to estimate the
power deviation $\Delta P$ of each comb line (see Fig. \ref{fig:AvoidedXssing}a).
The power deviation of the concerned comb lines evolves with the detuning,
abruptly transitioning to being enhanced or suppressed over a small
range of detuning, as illustrated in Figure \ref{fig:AvoidedXssing}b
. The deviations in the residual $\Delta P$ are detected, and reported
on Figure \ref{fig:AvoidedXssing}c. We observe here that the spectral
location of the mode crossing features in the comb spectrum is fixed
and match those of the modal deviations in the measured dispersion.
We also note a clear correlation between strongly enhanced comb lines
and the shift of the soliton centroid, that recoils away from these
lines. To further check the appearance of avoided mode crossings induced
recoil, we estimate the expected soliton recoil $\tilde{\Omega}$
based on the conservation of the spectral center of mass:
\begin{gather}
\int_{-\infty}^{+\infty}\mu\,A\,\sech^{2}\left(\dfrac{\mu\omega_{r}-\tilde{\Omega}}{B}\right)d\mu+\sum_{\mu}\mu\Delta P=0\nonumber \\
\Leftrightarrow\tilde{\Omega}=-\dfrac{\omega_{r}^{2}}{2AB}\sum_{\mu}\mu\Delta P,\label{eq:RecoilEstimate}
\end{gather}
This estimate is plotted in Figure \ref{fig:AvoidedXssing}e, together
with the fitted parameter $\Omega$ in \eqref{eq:FitExpr}, and an
overall agreement is found between these two values. It is interesting
to note that the soliton experiences a spectral recoil toward higher
optical frequencies, which is opposite to the so far reported frequency
shifts observed in microresonators in amorphous silica or silicon
nitride. Indeed, in these platforms, the first order Raman shock term
dominates and systemically shifts the frequency comb toward lower
frequencies, and can compensate the recoil induced by a dispersive
wave \cite{Karpov2016a,Yi2015}. The absence of Raman self-frequency
shift is expected in crystalline $\mgf$ platforms, where the Raman
gain is spectrally narrow \cite{Porto1967}.

The recoil on the soliton implies a change in the soliton's group
velocity and thus a modification of the comb repetition rate, according
to $\omega_{r}=D_{1}+\Omega\,D_{2}/D_{1}$ \cite{Yang2016c}, similar
to the Gordon–Haus effect in mode-locked lasers \cite{Paschotta2004,Haus1993}.
This is verified in Figure \ref{fig:AvoidedXssing}f, where the change
in the repetition rate frequency is plotted as a function of the measured
recoil and fitted with a linear model. The intercept matches the free
spectral range $D_{1}/2\pi$ and the slope yields $D_{2}/2\pi=1.72\:\pm0.48$~kHz,
which overlaps with the measured dispersion. The spread of the data-points
at small recoil values could originate from the thermal drift during
the measurement.

Overall, we observe that detuning dependent excitation of avoided
mode crossing are detrimental for the stability of the soliton Kerr
comb, and cause an enhanced sensitivity of the soliton repetition
rate to pump laser frequency fluctuations. At certain detuning points,
the excitation of mode crossings causes abrupt changes in the comb
repetition rate, resulting from the induced recoil, in agreement with
simulations performed in \cite{Matsko2014a}. The present method enables
the identification of detuning regions that minimize the impact of
avoided mode crossings. We also observed that the excitation of the
strong avoided mode crossing at $\delta\omega/2\pi=12$~MHz ($t\sim450$
s) is correlated with a sudden shift of the $\mathrm{\mathcal{S}}$-resonance
toward lower frequency (see VNA map Fig. \ref{fig:CombBW_detun}a).
This is not yet understood and will be investigated further in another
study.

At other detuning values $\delta\omega/2\pi=15.5,\ 15.9,\ 17.2$~MHz
($t\sim750,\ 780,\ 880$ s), the $\mathrm{\mathcal{S}}$-resonance
peak appears greatly enhanced. This is concomitant with the appearance
of sidebands around the repetition rate of the comb and of an amplitude
modulation of the soliton pulse train at a frequency of $\sim3.5$~MHz.
These observations suggest that the soliton is breathing, meaning
its amplitude and width oscillate in time, with a frequency typically
much smaller than the repetition rate \cite{Matsko2012,Yu2016,Lucas2016b,Bao2016a}.
While such instabilities are known to occur for small detuning values
\cite{Leo2013,Anderson:16}, they are unexpected for the large detuning
values explored in the frame of this work. Our experiments suggest
that the breathing of the soliton could be related and induced by
the mode crossing feature at $\mu=-106$. This observation of mode
crossing induced soliton breathing is reported here experimentally
for the first time and will be further investigated in a future work.

Nevertheless, it is interesting to point out that our observations
highlight the surprising robustness of the dissipative soliton, which
is sustained in the cavity in spite of all the reported perturbations.

\section{Discussion}

We demonstrated a novel technique to probe, stabilize, and control
the effective detuning of soliton states in optical microresonators
via a feedback on the pump laser frequency. It enables the experimental
study of the soliton's properties while varying the effective detuning
parameter and to verify the relation between this parameter and the
soliton duration. This relation is surprisingly well preserved although
the studied microresonator exhibits non-negligible deviations in its
mode spectrum in the form of avoided mode crossings. In addition,
we observed and studied the detuning-dependent mode crossing features
and associated spectral recoil that correlates with a modification
in the soliton round-trip time (repetition rate). These observations
of a detuning dependent repetition rate have important repercussions
for low phase noise microwave generation, as they enhance the transduction
of pump laser frequency noise onto noise in the soliton pulse repetition
rate. Furthermore, the mode crossings can also degrade the stability
of the soliton and induce breathing in a region where solitons are
expected to be stable.

Our method provides a way to experimentally explore the existence
range of the soliton and identify optimal sets of operating parameters
that favor a stable operation of the optical frequency comb. Moreover,
we reveal how these crossings induce deviations in the relation between
comb power and detuning, which can be a limitation for stabilization
techniques based on the comb power alone \cite{Yi2015,Yi2016}. The
presented method also enables the longterm operation of soliton-based
combs with stabilized bandwidth and power. The stabilization could
alternatively be achieved by direct actuation on the microresonator
\cite{Jost2015b,Joshi2016,Papp2013a}, to tune the free spectral range
and stabilize the cavity resonance on a stable pump laser. The fine
control of the two driving parameters of the nonlinear system (detuning
and pump power) will also enable the controlled access to various
soliton regimes predicted by the theory (soliton breathers, chaos)
\cite{Leo2013}. The presented observations could also provide insights
for sources of instabilities in systems described by the same type
of driven, damped Nonlinear Schrödinger Equation, such as rf-driven
waves in plasma \cite{Nozaki1986}, where similar probing and stabilization
schemes could be applied.

\bigskip

Note: during submission of this work, Yi, et al., reported on the
properties of single-mode dispersive waves induced by modal crossing
\cite{Yi2016b}.

\vspace*{\fill}
\begin{acknowledgments}
The authors acknowledge M. Gorodetsky for his fruitful discussions and suggestions, as well as M. Anderson, M. H. Pfeiffer, and V. Brasch for their valuable feedback on this paper. This work was supported by funding from the Swiss National Science Foundation under Grant Agreement No. 161573, as well as Contract No. W31P4Q-14-C-0050 (PULSE) from the Defense Advanced Research Projects Agency (DARPA), Defense Sciences Office (DSO). This material is based upon work supported by the Air Force Office of Scientific Research, Air Force Material Command, USAF under Award No. FA9550-15-1-0099. The authors acknowledge the support by the European Space Technology Centre with ESA Contracts No. 4000118777/16/NL/GM and No. 4000116145/16/NL/MH/GM, as well as funding from the European Union's FP7 programme under Marie Sklodowska-Curie ITN Grant Agreement No. 607493 and the European Union's Horizon 2020 research and innovation programme under Marie Sklodowska-Curie IF Grant Agreement No. 709249.
\end{acknowledgments}

\bibliographystyle{apsrev4-1}
\bibliography{Biblio_library}

\begin{thebibliography}{52}%
\makeatletter
\providecommand \@ifxundefined [1]{%
 \@ifx{#1\undefined}
}%
\providecommand \@ifnum [1]{%
 \ifnum #1\expandafter \@firstoftwo
 \else \expandafter \@secondoftwo
 \fi
}%
\providecommand \@ifx [1]{%
 \ifx #1\expandafter \@firstoftwo
 \else \expandafter \@secondoftwo
 \fi
}%
\providecommand \natexlab [1]{#1}%
\providecommand \enquote  [1]{``#1''}%
\providecommand \bibnamefont  [1]{#1}%
\providecommand \bibfnamefont [1]{#1}%
\providecommand \citenamefont [1]{#1}%
\providecommand \href@noop [0]{\@secondoftwo}%
\providecommand \href [0]{\begingroup \@sanitize@url \@href}%
\providecommand \@href[1]{\@@startlink{#1}\@@href}%
\providecommand \@@href[1]{\endgroup#1\@@endlink}%
\providecommand \@sanitize@url [0]{\catcode `\\12\catcode `\$12\catcode
  `\&12\catcode `\#12\catcode `\^12\catcode `\_12\catcode `\%12\relax}%
\providecommand \@@startlink[1]{}%
\providecommand \@@endlink[0]{}%
\providecommand \url  [0]{\begingroup\@sanitize@url \@url }%
\providecommand \@url [1]{\endgroup\@href {#1}{\urlprefix }}%
\providecommand \urlprefix  [0]{URL }%
\providecommand \Eprint [0]{\href }%
\providecommand \doibase [0]{http://dx.doi.org/}%
\providecommand \selectlanguage [0]{\@gobble}%
\providecommand \bibinfo  [0]{\@secondoftwo}%
\providecommand \bibfield  [0]{\@secondoftwo}%
\providecommand \translation [1]{[#1]}%
\providecommand \BibitemOpen [0]{}%
\providecommand \bibitemStop [0]{}%
\providecommand \bibitemNoStop [0]{.\EOS\space}%
\providecommand \EOS [0]{\spacefactor3000\relax}%
\providecommand \BibitemShut  [1]{\csname bibitem#1\endcsname}%
\let\auto@bib@innerbib\@empty
\bibitem [{\citenamefont {Leo}\ \emph {et~al.}(2010)\citenamefont {Leo},
  \citenamefont {Coen}, \citenamefont {Kockaert}, \citenamefont {Gorza},
  \citenamefont {Emplit},\ and\ \citenamefont {Haelterman}}]{Leo2010}%
  \BibitemOpen
  \bibfield  {author} {\bibinfo {author} {\bibfnamefont {F.}~\bibnamefont
  {Leo}}, \bibinfo {author} {\bibfnamefont {S.}~\bibnamefont {Coen}}, \bibinfo
  {author} {\bibfnamefont {P.}~\bibnamefont {Kockaert}}, \bibinfo {author}
  {\bibfnamefont {S.-P.}\ \bibnamefont {Gorza}}, \bibinfo {author}
  {\bibfnamefont {P.}~\bibnamefont {Emplit}}, \ and\ \bibinfo {author}
  {\bibfnamefont {M.}~\bibnamefont {Haelterman}},\ }\href {\doibase
  10.1038/NPHOTON.2010.120} {\bibfield  {journal} {\bibinfo  {journal} {Nature
  Photonics}\ }\textbf {\bibinfo {volume} {4}},\ \bibinfo {pages} {471}
  (\bibinfo {year} {2010})}\BibitemShut {NoStop}%
\bibitem [{\citenamefont {Herr}\ \emph {et~al.}(2013)\citenamefont {Herr},
  \citenamefont {Brasch}, \citenamefont {Jost}, \citenamefont {Wang},
  \citenamefont {Kondratiev}, \citenamefont {Gorodetsky},\ and\ \citenamefont
  {Kippenberg}}]{Herr2013}%
  \BibitemOpen
  \bibfield  {author} {\bibinfo {author} {\bibfnamefont {T.}~\bibnamefont
  {Herr}}, \bibinfo {author} {\bibfnamefont {V.}~\bibnamefont {Brasch}},
  \bibinfo {author} {\bibfnamefont {J.~D.}\ \bibnamefont {Jost}}, \bibinfo
  {author} {\bibfnamefont {C.~Y.}\ \bibnamefont {Wang}}, \bibinfo {author}
  {\bibfnamefont {N.~M.}\ \bibnamefont {Kondratiev}}, \bibinfo {author}
  {\bibfnamefont {M.~L.}\ \bibnamefont {Gorodetsky}}, \ and\ \bibinfo {author}
  {\bibfnamefont {T.~J.}\ \bibnamefont {Kippenberg}},\ }\href {\doibase
  10.1038/nphoton.2013.343} {\bibfield  {journal} {\bibinfo  {journal} {Nature
  Photonics}\ }\textbf {\bibinfo {volume} {8}},\ \bibinfo {pages} {145}
  (\bibinfo {year} {2013})}\BibitemShut {NoStop}%
\bibitem [{\citenamefont {Kippenberg}\ \emph {et~al.}(2011)\citenamefont
  {Kippenberg}, \citenamefont {Holzwarth},\ and\ \citenamefont
  {Diddams}}]{Kippenberg2011}%
  \BibitemOpen
  \bibfield  {author} {\bibinfo {author} {\bibfnamefont {T.~J.}\ \bibnamefont
  {Kippenberg}}, \bibinfo {author} {\bibfnamefont {R.}~\bibnamefont
  {Holzwarth}}, \ and\ \bibinfo {author} {\bibfnamefont {S.~A.}\ \bibnamefont
  {Diddams}},\ }\href {\doibase 10.1126/science.1193968} {\bibfield  {journal}
  {\bibinfo  {journal} {Science}\ }\textbf {\bibinfo {volume} {332}},\ \bibinfo
  {pages} {555} (\bibinfo {year} {2011})}\BibitemShut {NoStop}%
\bibitem [{\citenamefont {Del'Haye}\ \emph {et~al.}(2007)\citenamefont
  {Del'Haye}, \citenamefont {Schliesser}, \citenamefont {Arcizet},
  \citenamefont {Wilken}, \citenamefont {Holzwarth},\ and\ \citenamefont
  {Kippenberg}}]{Del'Haye2007}%
  \BibitemOpen
  \bibfield  {author} {\bibinfo {author} {\bibfnamefont {P.}~\bibnamefont
  {Del'Haye}}, \bibinfo {author} {\bibfnamefont {A.}~\bibnamefont
  {Schliesser}}, \bibinfo {author} {\bibfnamefont {O.}~\bibnamefont {Arcizet}},
  \bibinfo {author} {\bibfnamefont {T.}~\bibnamefont {Wilken}}, \bibinfo
  {author} {\bibfnamefont {R.}~\bibnamefont {Holzwarth}}, \ and\ \bibinfo
  {author} {\bibfnamefont {T.~J.}\ \bibnamefont {Kippenberg}},\ }\href
  {\doibase 10.1038/nature06401} {\bibfield  {journal} {\bibinfo  {journal}
  {Nature}\ }\textbf {\bibinfo {volume} {450}},\ \bibinfo {pages} {1214}
  (\bibinfo {year} {2007})}\BibitemShut {NoStop}%
\bibitem [{\citenamefont {Coen}\ \emph {et~al.}(2013)\citenamefont {Coen},
  \citenamefont {Randle}, \citenamefont {Sylvestre},\ and\ \citenamefont
  {Erkintalo}}]{Coen2013}%
  \BibitemOpen
  \bibfield  {author} {\bibinfo {author} {\bibfnamefont {S.}~\bibnamefont
  {Coen}}, \bibinfo {author} {\bibfnamefont {H.~G.}\ \bibnamefont {Randle}},
  \bibinfo {author} {\bibfnamefont {T.}~\bibnamefont {Sylvestre}}, \ and\
  \bibinfo {author} {\bibfnamefont {M.}~\bibnamefont {Erkintalo}},\ }\href
  {\doibase 10.1364/OL.38.000037} {\bibfield  {journal} {\bibinfo  {journal}
  {Optics letters}\ }\textbf {\bibinfo {volume} {38}},\ \bibinfo {pages} {37}
  (\bibinfo {year} {2013})}\BibitemShut {NoStop}%
\bibitem [{\citenamefont {Matsko}\ \emph {et~al.}(2011)\citenamefont {Matsko},
  \citenamefont {Savchenkov}, \citenamefont {Liang}, \citenamefont {Ilchenko},
  \citenamefont {Seidel},\ and\ \citenamefont {Maleki}}]{Matsko2011b}%
  \BibitemOpen
  \bibfield  {author} {\bibinfo {author} {\bibfnamefont {A.~B.}\ \bibnamefont
  {Matsko}}, \bibinfo {author} {\bibfnamefont {A.~A.}\ \bibnamefont
  {Savchenkov}}, \bibinfo {author} {\bibfnamefont {W.}~\bibnamefont {Liang}},
  \bibinfo {author} {\bibfnamefont {V.~S.}\ \bibnamefont {Ilchenko}}, \bibinfo
  {author} {\bibfnamefont {D.}~\bibnamefont {Seidel}}, \ and\ \bibinfo {author}
  {\bibfnamefont {L.}~\bibnamefont {Maleki}},\ }\href
  {http://www.ncbi.nlm.nih.gov/pubmed/21808332} {\bibfield  {journal} {\bibinfo
   {journal} {Optics letters}\ }\textbf {\bibinfo {volume} {36}},\ \bibinfo
  {pages} {2845} (\bibinfo {year} {2011})}\BibitemShut {NoStop}%
\bibitem [{\citenamefont {Akhmediev}\ and\ \citenamefont
  {Ankiewicz}(2003)}]{Akhmediev2003}%
  \BibitemOpen
  \bibfield  {author} {\bibinfo {author} {\bibfnamefont {N.~N.}\ \bibnamefont
  {Akhmediev}}\ and\ \bibinfo {author} {\bibfnamefont {A.}~\bibnamefont
  {Ankiewicz}},\ }\enquote {\bibinfo {title} {{Solitons Around Us: Integrable,
  Hamiltonian and Dissipative Systems}},}\ in\ \href {\doibase
  10.1007/3-540-36141-3_5} {\emph {\bibinfo {booktitle} {Optical Solitons:
  Theoretical and Experimental Challenges}}},\ \bibinfo {editor} {edited by\
  \bibinfo {editor} {\bibfnamefont {K.}~\bibnamefont {Porsezian}}\ and\
  \bibinfo {editor} {\bibfnamefont {V.~C.}\ \bibnamefont {Kuriakose}}}\
  (\bibinfo  {publisher} {Springer Berlin Heidelberg},\ \bibinfo {address}
  {Berlin, Heidelberg},\ \bibinfo {year} {2003})\ pp.\ \bibinfo {pages}
  {105--126}\BibitemShut {NoStop}%
\bibitem [{\citenamefont {Brasch}\ \emph {et~al.}(2015)\citenamefont {Brasch},
  \citenamefont {Geiselmann}, \citenamefont {Herr}, \citenamefont {Lihachev},
  \citenamefont {Pfeiffer}, \citenamefont {Gorodetsky},\ and\ \citenamefont
  {Kippenberg}}]{Brasch2015}%
  \BibitemOpen
  \bibfield  {author} {\bibinfo {author} {\bibfnamefont {V.}~\bibnamefont
  {Brasch}}, \bibinfo {author} {\bibfnamefont {M.}~\bibnamefont {Geiselmann}},
  \bibinfo {author} {\bibfnamefont {T.}~\bibnamefont {Herr}}, \bibinfo {author}
  {\bibfnamefont {G.}~\bibnamefont {Lihachev}}, \bibinfo {author}
  {\bibfnamefont {M.~H.~P.}\ \bibnamefont {Pfeiffer}}, \bibinfo {author}
  {\bibfnamefont {M.~L.}\ \bibnamefont {Gorodetsky}}, \ and\ \bibinfo {author}
  {\bibfnamefont {T.~J.}\ \bibnamefont {Kippenberg}},\ }\href {\doibase
  10.1126/science.aad4811} {\bibfield  {journal} {\bibinfo  {journal}
  {Science}\ }\textbf {\bibinfo {volume} {351}},\ \bibinfo {pages} {357}
  (\bibinfo {year} {2015})}\BibitemShut {NoStop}%
\bibitem [{\citenamefont {Yi}\ \emph {et~al.}(2015)\citenamefont {Yi},
  \citenamefont {Yang}, \citenamefont {Yang}, \citenamefont {Suh},\ and\
  \citenamefont {Vahala}}]{Yi2015}%
  \BibitemOpen
  \bibfield  {author} {\bibinfo {author} {\bibfnamefont {X.}~\bibnamefont
  {Yi}}, \bibinfo {author} {\bibfnamefont {Q.-F.}\ \bibnamefont {Yang}},
  \bibinfo {author} {\bibfnamefont {K.~Y.}\ \bibnamefont {Yang}}, \bibinfo
  {author} {\bibfnamefont {M.-G.}\ \bibnamefont {Suh}}, \ and\ \bibinfo
  {author} {\bibfnamefont {K.}~\bibnamefont {Vahala}},\ }\href {\doibase
  10.1364/OPTICA.2.001078} {\bibfield  {journal} {\bibinfo  {journal} {Optica}\
  }\textbf {\bibinfo {volume} {2}},\ \bibinfo {pages} {1078} (\bibinfo {year}
  {2015})}\BibitemShut {NoStop}%
\bibitem [{\citenamefont {Joshi}\ \emph {et~al.}(2016)\citenamefont {Joshi},
  \citenamefont {Jang}, \citenamefont {Luke}, \citenamefont {Ji}, \citenamefont
  {Miller}, \citenamefont {Klenner}, \citenamefont {Okawachi}, \citenamefont
  {Lipson},\ and\ \citenamefont {Gaeta}}]{Joshi2016}%
  \BibitemOpen
  \bibfield  {author} {\bibinfo {author} {\bibfnamefont {C.}~\bibnamefont
  {Joshi}}, \bibinfo {author} {\bibfnamefont {J.~K.}\ \bibnamefont {Jang}},
  \bibinfo {author} {\bibfnamefont {K.}~\bibnamefont {Luke}}, \bibinfo {author}
  {\bibfnamefont {X.}~\bibnamefont {Ji}}, \bibinfo {author} {\bibfnamefont
  {S.~A.}\ \bibnamefont {Miller}}, \bibinfo {author} {\bibfnamefont
  {A.}~\bibnamefont {Klenner}}, \bibinfo {author} {\bibfnamefont
  {Y.}~\bibnamefont {Okawachi}}, \bibinfo {author} {\bibfnamefont
  {M.}~\bibnamefont {Lipson}}, \ and\ \bibinfo {author} {\bibfnamefont {A.~L.}\
  \bibnamefont {Gaeta}},\ }\href {\doibase 10.1364/OL.41.002565} {\bibfield
  {journal} {\bibinfo  {journal} {Optics Letters}\ }\textbf {\bibinfo {volume}
  {41}},\ \bibinfo {pages} {2565} (\bibinfo {year} {2016})}\BibitemShut
  {NoStop}%
\bibitem [{\citenamefont {Marin-Palomo}\ \emph {et~al.}(2016)\citenamefont
  {Marin-Palomo}, \citenamefont {Kemal}, \citenamefont {Karpov}, \citenamefont
  {Kordts}, \citenamefont {Pfeifle}, \citenamefont {Pfeiffer}, \citenamefont
  {Trocha}, \citenamefont {Wolf}, \citenamefont {Brasch}, \citenamefont
  {Rosenberger}, \citenamefont {Vijayan}, \citenamefont {Freude}, \citenamefont
  {Kippenberg},\ and\ \citenamefont {Koos}}]{Marin-Palomo2016}%
  \BibitemOpen
  \bibfield  {author} {\bibinfo {author} {\bibfnamefont {P.}~\bibnamefont
  {Marin-Palomo}}, \bibinfo {author} {\bibfnamefont {J.~N.}\ \bibnamefont
  {Kemal}}, \bibinfo {author} {\bibfnamefont {M.}~\bibnamefont {Karpov}},
  \bibinfo {author} {\bibfnamefont {A.}~\bibnamefont {Kordts}}, \bibinfo
  {author} {\bibfnamefont {J.}~\bibnamefont {Pfeifle}}, \bibinfo {author}
  {\bibfnamefont {M.~H.~P.}\ \bibnamefont {Pfeiffer}}, \bibinfo {author}
  {\bibfnamefont {P.}~\bibnamefont {Trocha}}, \bibinfo {author} {\bibfnamefont
  {S.}~\bibnamefont {Wolf}}, \bibinfo {author} {\bibfnamefont {V.}~\bibnamefont
  {Brasch}}, \bibinfo {author} {\bibfnamefont {R.}~\bibnamefont {Rosenberger}},
  \bibinfo {author} {\bibfnamefont {K.}~\bibnamefont {Vijayan}}, \bibinfo
  {author} {\bibfnamefont {W.}~\bibnamefont {Freude}}, \bibinfo {author}
  {\bibfnamefont {T.~J.}\ \bibnamefont {Kippenberg}}, \ and\ \bibinfo {author}
  {\bibfnamefont {C.}~\bibnamefont {Koos}},\ }\href
  {http://arxiv.org/abs/1610.01484} {\ ,\ \bibinfo {pages} {13} (\bibinfo
  {year} {2016})},\ \Eprint {http://arxiv.org/abs/1610.01484}
  {arXiv:1610.01484} \BibitemShut {NoStop}%
\bibitem [{\citenamefont {Jost}\ \emph
  {et~al.}(2015{\natexlab{a}})\citenamefont {Jost}, \citenamefont {Herr},
  \citenamefont {Lecaplain}, \citenamefont {Brasch}, \citenamefont {Pfeiffer},\
  and\ \citenamefont {Kippenberg}}]{Jost15}%
  \BibitemOpen
  \bibfield  {author} {\bibinfo {author} {\bibfnamefont {J.~D.}\ \bibnamefont
  {Jost}}, \bibinfo {author} {\bibfnamefont {T.}~\bibnamefont {Herr}}, \bibinfo
  {author} {\bibfnamefont {C.}~\bibnamefont {Lecaplain}}, \bibinfo {author}
  {\bibfnamefont {V.}~\bibnamefont {Brasch}}, \bibinfo {author} {\bibfnamefont
  {M.~H.~P.}\ \bibnamefont {Pfeiffer}}, \ and\ \bibinfo {author} {\bibfnamefont
  {T.~J.}\ \bibnamefont {Kippenberg}},\ }\href {\doibase
  10.1364/OPTICA.2.000706} {\bibfield  {journal} {\bibinfo  {journal} {Optica}\
  }\textbf {\bibinfo {volume} {2}},\ \bibinfo {pages} {706} (\bibinfo {year}
  {2015}{\natexlab{a}})}\BibitemShut {NoStop}%
\bibitem [{\citenamefont {Del'Haye}\ \emph {et~al.}(2016)\citenamefont
  {Del'Haye}, \citenamefont {Coillet}, \citenamefont {Fortier}, \citenamefont
  {Beha}, \citenamefont {Cole}, \citenamefont {Yang}, \citenamefont {Lee},
  \citenamefont {Vahala}, \citenamefont {Papp},\ and\ \citenamefont
  {Diddams}}]{DelHaye2016}%
  \BibitemOpen
  \bibfield  {author} {\bibinfo {author} {\bibfnamefont {P.}~\bibnamefont
  {Del'Haye}}, \bibinfo {author} {\bibfnamefont {A.}~\bibnamefont {Coillet}},
  \bibinfo {author} {\bibfnamefont {T.}~\bibnamefont {Fortier}}, \bibinfo
  {author} {\bibfnamefont {K.}~\bibnamefont {Beha}}, \bibinfo {author}
  {\bibfnamefont {D.~C.}\ \bibnamefont {Cole}}, \bibinfo {author}
  {\bibfnamefont {K.~Y.}\ \bibnamefont {Yang}}, \bibinfo {author}
  {\bibfnamefont {H.}~\bibnamefont {Lee}}, \bibinfo {author} {\bibfnamefont
  {K.~J.}\ \bibnamefont {Vahala}}, \bibinfo {author} {\bibfnamefont {S.~B.}\
  \bibnamefont {Papp}}, \ and\ \bibinfo {author} {\bibfnamefont {S.~A.}\
  \bibnamefont {Diddams}},\ }\href {\doibase 10.1038/nphoton.2016.105}
  {\bibfield  {journal} {\bibinfo  {journal} {Nature Photonics}\ }\textbf
  {\bibinfo {volume} {10}},\ \bibinfo {pages} {1} (\bibinfo {year}
  {2016})}\BibitemShut {NoStop}%
\bibitem [{\citenamefont {Brasch}\ \emph {et~al.}(2017)\citenamefont {Brasch},
  \citenamefont {Lucas}, \citenamefont {Jost}, \citenamefont {Geiselmann},\
  and\ \citenamefont {Kippenberg}}]{Brasch2017}%
  \BibitemOpen
  \bibfield  {author} {\bibinfo {author} {\bibfnamefont {V.}~\bibnamefont
  {Brasch}}, \bibinfo {author} {\bibfnamefont {E.}~\bibnamefont {Lucas}},
  \bibinfo {author} {\bibfnamefont {J.~D.}\ \bibnamefont {Jost}}, \bibinfo
  {author} {\bibfnamefont {M.}~\bibnamefont {Geiselmann}}, \ and\ \bibinfo
  {author} {\bibfnamefont {T.~J.}\ \bibnamefont {Kippenberg}},\ }\href
  {\doibase 10.1038/lsa.2016.202} {\bibfield  {journal} {\bibinfo  {journal}
  {Light Sci Appl.}\ }\textbf {\bibinfo {volume} {6}},\ \bibinfo {pages}
  {e16202} (\bibinfo {year} {2017})}\BibitemShut {NoStop}%
\bibitem [{\citenamefont {Liang}\ \emph {et~al.}(2015)\citenamefont {Liang},
  \citenamefont {Ilchenko}, \citenamefont {Eliyahu}, \citenamefont
  {Savchenkov}, \citenamefont {Matsko}, \citenamefont {Seidel},\ and\
  \citenamefont {Maleki}}]{Liang2015a}%
  \BibitemOpen
  \bibfield  {author} {\bibinfo {author} {\bibfnamefont {W.}~\bibnamefont
  {Liang}}, \bibinfo {author} {\bibfnamefont {V.~S.}\ \bibnamefont {Ilchenko}},
  \bibinfo {author} {\bibfnamefont {D.}~\bibnamefont {Eliyahu}}, \bibinfo
  {author} {\bibfnamefont {a.~a.}\ \bibnamefont {Savchenkov}}, \bibinfo
  {author} {\bibfnamefont {a.~B.}\ \bibnamefont {Matsko}}, \bibinfo {author}
  {\bibfnamefont {D.}~\bibnamefont {Seidel}}, \ and\ \bibinfo {author}
  {\bibfnamefont {L.}~\bibnamefont {Maleki}},\ }\href {\doibase
  10.1038/ncomms8371} {\bibfield  {journal} {\bibinfo  {journal} {Nature
  Communications}\ }\textbf {\bibinfo {volume} {6}},\ \bibinfo {pages} {7371}
  (\bibinfo {year} {2015})}\BibitemShut {NoStop}%
\bibitem [{\citenamefont {Jang}\ \emph {et~al.}(2014)\citenamefont {Jang},
  \citenamefont {Erkintalo}, \citenamefont {Murdoch},\ and\ \citenamefont
  {Coen}}]{Jang2014}%
  \BibitemOpen
  \bibfield  {author} {\bibinfo {author} {\bibfnamefont {J.~K.}\ \bibnamefont
  {Jang}}, \bibinfo {author} {\bibfnamefont {M.}~\bibnamefont {Erkintalo}},
  \bibinfo {author} {\bibfnamefont {S.~G.}\ \bibnamefont {Murdoch}}, \ and\
  \bibinfo {author} {\bibfnamefont {S.~S.}\ \bibnamefont {Coen}},\ }\href
  {\doibase 10.1364/OL.39.005503} {\bibfield  {journal} {\bibinfo  {journal}
  {Optics Letters}\ }\textbf {\bibinfo {volume} {39}},\ \bibinfo {pages} {5503}
  (\bibinfo {year} {2014})}\BibitemShut {NoStop}%
\bibitem [{\citenamefont {Akhmediev}\ and\ \citenamefont
  {Karlsson}(1995)}]{Akhmediev1995a}%
  \BibitemOpen
  \bibfield  {author} {\bibinfo {author} {\bibfnamefont {N.}~\bibnamefont
  {Akhmediev}}\ and\ \bibinfo {author} {\bibfnamefont {M.}~\bibnamefont
  {Karlsson}},\ }\href {\doibase 10.1103/PhysRevA.51.2602} {\bibfield
  {journal} {\bibinfo  {journal} {Physical Review A}\ }\textbf {\bibinfo
  {volume} {51}},\ \bibinfo {pages} {2602} (\bibinfo {year}
  {1995})}\BibitemShut {NoStop}%
\bibitem [{\citenamefont {Coen}\ and\ \citenamefont
  {Erkintalo}(2013)}]{Coen2013a}%
  \BibitemOpen
  \bibfield  {author} {\bibinfo {author} {\bibfnamefont {S.}~\bibnamefont
  {Coen}}\ and\ \bibinfo {author} {\bibfnamefont {M.}~\bibnamefont
  {Erkintalo}},\ }\href {\doibase 10.1364/OL.38.001790} {\bibfield  {journal}
  {\bibinfo  {journal} {Optics letters}\ }\textbf {\bibinfo {volume} {38}},\
  \bibinfo {pages} {1790} (\bibinfo {year} {2013})}\BibitemShut {NoStop}%
\bibitem [{\citenamefont {Carmon}\ \emph {et~al.}(2004)\citenamefont {Carmon},
  \citenamefont {Yang},\ and\ \citenamefont {Vahala}}]{Carmon2004}%
  \BibitemOpen
  \bibfield  {author} {\bibinfo {author} {\bibfnamefont {T.}~\bibnamefont
  {Carmon}}, \bibinfo {author} {\bibfnamefont {L.}~\bibnamefont {Yang}}, \ and\
  \bibinfo {author} {\bibfnamefont {K.~J.}\ \bibnamefont {Vahala}},\ }\href
  {\doibase 10.1364/OPEX.12.004742} {\bibfield  {journal} {\bibinfo  {journal}
  {Optics Express}\ }\textbf {\bibinfo {volume} {12}},\ \bibinfo {pages} {4742}
  (\bibinfo {year} {2004})}\BibitemShut {NoStop}%
\bibitem [{\citenamefont {Del'Haye}\ \emph {et~al.}(2015)\citenamefont
  {Del'Haye}, \citenamefont {Coillet}, \citenamefont {Loh}, \citenamefont
  {Beha}, \citenamefont {Papp}, \citenamefont {Diddams}, \citenamefont
  {Del'Haye}, \citenamefont {Coillet}, \citenamefont {Loh}, \citenamefont
  {Beha}, \citenamefont {Papp},\ and\ \citenamefont {Diddams}}]{DelHaye2015a}%
  \BibitemOpen
  \bibfield  {author} {\bibinfo {author} {\bibfnamefont {P.}~\bibnamefont
  {Del'Haye}}, \bibinfo {author} {\bibfnamefont {A.}~\bibnamefont {Coillet}},
  \bibinfo {author} {\bibfnamefont {W.}~\bibnamefont {Loh}}, \bibinfo {author}
  {\bibfnamefont {K.}~\bibnamefont {Beha}}, \bibinfo {author} {\bibfnamefont
  {S.~B.}\ \bibnamefont {Papp}}, \bibinfo {author} {\bibfnamefont {S.~A.}\
  \bibnamefont {Diddams}}, \bibinfo {author} {\bibfnamefont {P.}~\bibnamefont
  {Del'Haye}}, \bibinfo {author} {\bibfnamefont {A.}~\bibnamefont {Coillet}},
  \bibinfo {author} {\bibfnamefont {W.}~\bibnamefont {Loh}}, \bibinfo {author}
  {\bibfnamefont {K.}~\bibnamefont {Beha}}, \bibinfo {author} {\bibfnamefont
  {S.~B.}\ \bibnamefont {Papp}}, \ and\ \bibinfo {author} {\bibfnamefont
  {S.~A.}\ \bibnamefont {Diddams}},\ }\href {\doibase 10.1038/ncomms6668}
  {\bibfield  {journal} {\bibinfo  {journal} {Nature Communications}\ }\textbf
  {\bibinfo {volume} {6}},\ \bibinfo {pages} {1} (\bibinfo {year}
  {2015})}\BibitemShut {NoStop}%
\bibitem [{\citenamefont {Matsko}\ and\ \citenamefont
  {Maleki}(2015)}]{Matsko2015}%
  \BibitemOpen
  \bibfield  {author} {\bibinfo {author} {\bibfnamefont {A.~B.}\ \bibnamefont
  {Matsko}}\ and\ \bibinfo {author} {\bibfnamefont {L.}~\bibnamefont
  {Maleki}},\ }\href {\doibase 10.1103/PhysRevA.91.013831} {\bibfield
  {journal} {\bibinfo  {journal} {Physical Review A}\ }\textbf {\bibinfo
  {volume} {91}},\ \bibinfo {pages} {013831} (\bibinfo {year}
  {2015})}\BibitemShut {NoStop}%
\bibitem [{\citenamefont {Guo}\ \emph {et~al.}(2017)\citenamefont {Guo},
  \citenamefont {Karpov}, \citenamefont {Lucas}, \citenamefont {Kordts},
  \citenamefont {Pfeiffer}, \citenamefont {Brasch}, \citenamefont {Lihachev},
  \citenamefont {Lobanov}, \citenamefont {Gorodetsky},\ and\ \citenamefont
  {Kippenberg}}]{Guo2016}%
  \BibitemOpen
  \bibfield  {author} {\bibinfo {author} {\bibfnamefont {H.}~\bibnamefont
  {Guo}}, \bibinfo {author} {\bibfnamefont {M.}~\bibnamefont {Karpov}},
  \bibinfo {author} {\bibfnamefont {E.}~\bibnamefont {Lucas}}, \bibinfo
  {author} {\bibfnamefont {A.}~\bibnamefont {Kordts}}, \bibinfo {author}
  {\bibfnamefont {M.~H.~P.}\ \bibnamefont {Pfeiffer}}, \bibinfo {author}
  {\bibfnamefont {V.}~\bibnamefont {Brasch}}, \bibinfo {author} {\bibfnamefont
  {G.}~\bibnamefont {Lihachev}}, \bibinfo {author} {\bibfnamefont {V.~E.}\
  \bibnamefont {Lobanov}}, \bibinfo {author} {\bibfnamefont {M.~L.}\
  \bibnamefont {Gorodetsky}}, \ and\ \bibinfo {author} {\bibfnamefont {T.~J.}\
  \bibnamefont {Kippenberg}},\ }\href {http://dx.doi.org/10.1038/nphys3893
  http://10.0.4.14/nphys3893
  http://www.nature.com/nphys/journal/vaop/ncurrent/abs/nphys3893.html{\#}supplementary-information
  http://www.nature.com/nphys/journal/v13/n1/abs/nphys3893.html{\#}supplementary-information}
  {\bibfield  {journal} {\bibinfo  {journal} {Nature Physics}\ }\textbf
  {\bibinfo {volume} {13}},\ \bibinfo {pages} {94} (\bibinfo {year}
  {2017})}\BibitemShut {NoStop}%
\bibitem [{\citenamefont {Herr}\ \emph {et~al.}(2012)\citenamefont {Herr},
  \citenamefont {Hartinger}, \citenamefont {Riemensberger}, \citenamefont
  {Wang}, \citenamefont {Gavartin}, \citenamefont {Holzwarth}, \citenamefont
  {Gorodetsky},\ and\ \citenamefont {Kippenberg}}]{Herr2012}%
  \BibitemOpen
  \bibfield  {author} {\bibinfo {author} {\bibfnamefont {T.}~\bibnamefont
  {Herr}}, \bibinfo {author} {\bibfnamefont {K.}~\bibnamefont {Hartinger}},
  \bibinfo {author} {\bibfnamefont {J.}~\bibnamefont {Riemensberger}}, \bibinfo
  {author} {\bibfnamefont {C.~Y.}\ \bibnamefont {Wang}}, \bibinfo {author}
  {\bibfnamefont {E.}~\bibnamefont {Gavartin}}, \bibinfo {author}
  {\bibfnamefont {R.}~\bibnamefont {Holzwarth}}, \bibinfo {author}
  {\bibfnamefont {M.~L.}\ \bibnamefont {Gorodetsky}}, \ and\ \bibinfo {author}
  {\bibfnamefont {T.~J.}\ \bibnamefont {Kippenberg}},\ }\href {\doibase
  10.1038/nphoton.2012.127} {\bibfield  {journal} {\bibinfo  {journal} {Nature
  Photonics}\ }\textbf {\bibinfo {volume} {6}},\ \bibinfo {pages} {480}
  (\bibinfo {year} {2012})}\BibitemShut {NoStop}%
\bibitem [{\citenamefont {Chembo}\ and\ \citenamefont
  {Menyuk}(2013)}]{Chembo2013}%
  \BibitemOpen
  \bibfield  {author} {\bibinfo {author} {\bibfnamefont {Y.~K.}\ \bibnamefont
  {Chembo}}\ and\ \bibinfo {author} {\bibfnamefont {C.~R.}\ \bibnamefont
  {Menyuk}},\ }\href {\doibase 10.1103/PhysRevA.87.053852} {\bibfield
  {journal} {\bibinfo  {journal} {Phys. Rev. A}\ }\textbf {\bibinfo {volume}
  {87}},\ \bibinfo {pages} {053852} (\bibinfo {year} {2013})}\BibitemShut
  {NoStop}%
\bibitem [{\citenamefont {Wabnitz}(1993)}]{Wabnitz1993}%
  \BibitemOpen
  \bibfield  {author} {\bibinfo {author} {\bibfnamefont {S.}~\bibnamefont
  {Wabnitz}},\ }\href {http://www.ncbi.nlm.nih.gov/pubmed/19802213} {\bibfield
  {journal} {\bibinfo  {journal} {Opt. Lett.}\ }\textbf {\bibinfo {volume}
  {18}},\ \bibinfo {pages} {601} (\bibinfo {year} {1993})}\BibitemShut
  {NoStop}%
\bibitem [{\citenamefont {Lugiato}\ and\ \citenamefont
  {Lefever}(1987)}]{Lugiato1987}%
  \BibitemOpen
  \bibfield  {author} {\bibinfo {author} {\bibfnamefont {L.~A.}\ \bibnamefont
  {Lugiato}}\ and\ \bibinfo {author} {\bibfnamefont {R.}~\bibnamefont
  {Lefever}},\ }\href
  {http://journals.aps.org/prl/abstract/10.1103/PhysRevLett.58.2209} {\bibfield
   {journal} {\bibinfo  {journal} {Physical review letters}\ }\textbf {\bibinfo
  {volume} {58}},\ \bibinfo {pages} {2209} (\bibinfo {year}
  {1987})}\BibitemShut {NoStop}%
\bibitem [{\citenamefont {Wahlstrand}\ \emph {et~al.}(2007)\citenamefont
  {Wahlstrand}, \citenamefont {Willits}, \citenamefont {Schibli}, \citenamefont
  {Menyuk},\ and\ \citenamefont {Cundiff}}]{Wahlstrand2007}%
  \BibitemOpen
  \bibfield  {author} {\bibinfo {author} {\bibfnamefont {J.~K.}\ \bibnamefont
  {Wahlstrand}}, \bibinfo {author} {\bibfnamefont {J.~T.}\ \bibnamefont
  {Willits}}, \bibinfo {author} {\bibfnamefont {T.~R.}\ \bibnamefont
  {Schibli}}, \bibinfo {author} {\bibfnamefont {C.~R.}\ \bibnamefont {Menyuk}},
  \ and\ \bibinfo {author} {\bibfnamefont {S.~T.}\ \bibnamefont {Cundiff}},\
  }\href {\doibase 10.1364/OL.32.003426} {\bibfield  {journal} {\bibinfo
  {journal} {Optics Letters}\ }\textbf {\bibinfo {volume} {32}},\ \bibinfo
  {pages} {3426} (\bibinfo {year} {2007})}\BibitemShut {NoStop}%
\bibitem [{\citenamefont {Lee}\ and\ \citenamefont {Schibli}(2014)}]{Lee2014}%
  \BibitemOpen
  \bibfield  {author} {\bibinfo {author} {\bibfnamefont {C.~C.}\ \bibnamefont
  {Lee}}\ and\ \bibinfo {author} {\bibfnamefont {T.~R.}\ \bibnamefont
  {Schibli}},\ }\href {\doibase 10.1103/PhysRevLett.112.223903} {\bibfield
  {journal} {\bibinfo  {journal} {Physical Review Letters}\ }\textbf {\bibinfo
  {volume} {112}},\ \bibinfo {pages} {223903} (\bibinfo {year}
  {2014})}\BibitemShut {NoStop}%
\bibitem [{\citenamefont {Herr}\ \emph {et~al.}(2014)\citenamefont {Herr},
  \citenamefont {Brasch}, \citenamefont {Jost}, \citenamefont {Mirgorodskiy},
  \citenamefont {Lihachev}, \citenamefont {Gorodetsky},\ and\ \citenamefont
  {Kippenberg}}]{Herr2013b}%
  \BibitemOpen
  \bibfield  {author} {\bibinfo {author} {\bibfnamefont {T.}~\bibnamefont
  {Herr}}, \bibinfo {author} {\bibfnamefont {V.}~\bibnamefont {Brasch}},
  \bibinfo {author} {\bibfnamefont {J.~D.}\ \bibnamefont {Jost}}, \bibinfo
  {author} {\bibfnamefont {I.}~\bibnamefont {Mirgorodskiy}}, \bibinfo {author}
  {\bibfnamefont {G.}~\bibnamefont {Lihachev}}, \bibinfo {author}
  {\bibfnamefont {M.~L.}\ \bibnamefont {Gorodetsky}}, \ and\ \bibinfo {author}
  {\bibfnamefont {T.~J.}\ \bibnamefont {Kippenberg}},\ }\href {\doibase
  10.1103/PhysRevLett.113.123901} {\bibfield  {journal} {\bibinfo  {journal}
  {Physical Review Letters}\ }\textbf {\bibinfo {volume} {113}},\ \bibinfo
  {pages} {123901} (\bibinfo {year} {2014})}\BibitemShut {NoStop}%
\bibitem [{\citenamefont {Del'Haye}\ \emph {et~al.}(2009)\citenamefont
  {Del'Haye}, \citenamefont {Arcizet}, \citenamefont {Gorodetsky},
  \citenamefont {Holzwarth},\ and\ \citenamefont {Kippenberg}}]{Del'Haye2009}%
  \BibitemOpen
  \bibfield  {author} {\bibinfo {author} {\bibfnamefont {P.}~\bibnamefont
  {Del'Haye}}, \bibinfo {author} {\bibfnamefont {O.}~\bibnamefont {Arcizet}},
  \bibinfo {author} {\bibfnamefont {M.~L.}\ \bibnamefont {Gorodetsky}},
  \bibinfo {author} {\bibfnamefont {R.}~\bibnamefont {Holzwarth}}, \ and\
  \bibinfo {author} {\bibfnamefont {T.~J.}\ \bibnamefont {Kippenberg}},\ }\href
  {\doibase 10.1038/nphoton.2009.138} {\bibfield  {journal} {\bibinfo
  {journal} {Nature Photonics}\ }\textbf {\bibinfo {volume} {3}},\ \bibinfo
  {pages} {529} (\bibinfo {year} {2009})}\BibitemShut {NoStop}%
\bibitem [{\citenamefont {Grudinin}\ \emph {et~al.}(2013)\citenamefont
  {Grudinin}, \citenamefont {Baumgartel},\ and\ \citenamefont
  {Yu}}]{Grudinin2013b}%
  \BibitemOpen
  \bibfield  {author} {\bibinfo {author} {\bibfnamefont {I.~S.}\ \bibnamefont
  {Grudinin}}, \bibinfo {author} {\bibfnamefont {L.}~\bibnamefont
  {Baumgartel}}, \ and\ \bibinfo {author} {\bibfnamefont {N.}~\bibnamefont
  {Yu}},\ }\href {\doibase 10.1364/OE.21.026929} {\bibfield  {journal}
  {\bibinfo  {journal} {Optics express}\ }\textbf {\bibinfo {volume} {21}},\
  \bibinfo {pages} {26929} (\bibinfo {year} {2013})}\BibitemShut {NoStop}%
\bibitem [{\citenamefont {Zhou}\ \emph {et~al.}(2015)\citenamefont {Zhou},
  \citenamefont {Huang}, \citenamefont {Dong}, \citenamefont {Liao},
  \citenamefont {Qiu},\ and\ \citenamefont {Wong}}]{Zhou2015}%
  \BibitemOpen
  \bibfield  {author} {\bibinfo {author} {\bibfnamefont {H.}~\bibnamefont
  {Zhou}}, \bibinfo {author} {\bibfnamefont {S.~W.}\ \bibnamefont {Huang}},
  \bibinfo {author} {\bibfnamefont {Y.}~\bibnamefont {Dong}}, \bibinfo {author}
  {\bibfnamefont {M.}~\bibnamefont {Liao}}, \bibinfo {author} {\bibfnamefont
  {K.}~\bibnamefont {Qiu}}, \ and\ \bibinfo {author} {\bibfnamefont {C.~W.}\
  \bibnamefont {Wong}},\ }\href {\doibase 10.1109/JPHOT.2015.2433902}
  {\bibfield  {journal} {\bibinfo  {journal} {IEEE Photonics Journal}\ }\textbf
  {\bibinfo {volume} {7}},\ \bibinfo {pages} {1} (\bibinfo {year}
  {2015})}\BibitemShut {NoStop}%
\bibitem [{\citenamefont {Mili{\'{a}}n}\ and\ \citenamefont
  {Skryabin}(2014)}]{Milian2014}%
  \BibitemOpen
  \bibfield  {author} {\bibinfo {author} {\bibfnamefont {C.}~\bibnamefont
  {Mili{\'{a}}n}}\ and\ \bibinfo {author} {\bibfnamefont {D.~V.}\ \bibnamefont
  {Skryabin}},\ }\href {\doibase 10.1364/OE.22.003732} {\bibfield  {journal}
  {\bibinfo  {journal} {Optics Express}\ }\textbf {\bibinfo {volume} {22}},\
  \bibinfo {pages} {3732} (\bibinfo {year} {2014})}\BibitemShut {NoStop}%
\bibitem [{\citenamefont {Matsko}\ \emph {et~al.}(2016)\citenamefont {Matsko},
  \citenamefont {Liang}, \citenamefont {Savchenkov}, \citenamefont {Eliyahu},\
  and\ \citenamefont {Maleki}}]{Matsko2016a}%
  \BibitemOpen
  \bibfield  {author} {\bibinfo {author} {\bibfnamefont {A.~B.}\ \bibnamefont
  {Matsko}}, \bibinfo {author} {\bibfnamefont {W.}~\bibnamefont {Liang}},
  \bibinfo {author} {\bibfnamefont {A.~A.}\ \bibnamefont {Savchenkov}},
  \bibinfo {author} {\bibfnamefont {D.}~\bibnamefont {Eliyahu}}, \ and\
  \bibinfo {author} {\bibfnamefont {L.}~\bibnamefont {Maleki}},\ }\href
  {\doibase 10.1364/OL.41.002907} {\bibfield  {journal} {\bibinfo  {journal}
  {Optics Letters}\ }\textbf {\bibinfo {volume} {41}},\ \bibinfo {pages} {2907}
  (\bibinfo {year} {2016})}\BibitemShut {NoStop}%
\bibitem [{\citenamefont {Wang}\ \emph {et~al.}(2016)\citenamefont {Wang},
  \citenamefont {Leo}, \citenamefont {Fatome}, \citenamefont {Luo},
  \citenamefont {Jang}, \citenamefont {Erkintalo}, \citenamefont {Murdoch},\
  and\ \citenamefont {Coen}}]{Wang2016}%
  \BibitemOpen
  \bibfield  {author} {\bibinfo {author} {\bibfnamefont {Y.}~\bibnamefont
  {Wang}}, \bibinfo {author} {\bibfnamefont {F.}~\bibnamefont {Leo}}, \bibinfo
  {author} {\bibfnamefont {J.}~\bibnamefont {Fatome}}, \bibinfo {author}
  {\bibfnamefont {K.}~\bibnamefont {Luo}}, \bibinfo {author} {\bibfnamefont
  {J.~K.}\ \bibnamefont {Jang}}, \bibinfo {author} {\bibfnamefont {M.~J.}\
  \bibnamefont {Erkintalo}}, \bibinfo {author} {\bibfnamefont {S.~G.}\
  \bibnamefont {Murdoch}}, \ and\ \bibinfo {author} {\bibfnamefont
  {S.}~\bibnamefont {Coen}},\ }in\ \href {\doibase
  10.1364/CLEO_QELS.2016.FF2A.6} {\emph {\bibinfo {booktitle} {Conference on
  Lasers and Electro-Optics}}},\ \bibinfo {series and number} {OSA Technical
  Digest (online)}\ (\bibinfo  {publisher} {Optical Society of America},\
  \bibinfo {address} {San Jose, California},\ \bibinfo {year} {2016})\ p.\
  \bibinfo {pages} {FF2A.6}\BibitemShut {NoStop}%
\bibitem [{\citenamefont {Karpov}\ \emph {et~al.}(2016)\citenamefont {Karpov},
  \citenamefont {Guo}, \citenamefont {Kordts}, \citenamefont {Brasch},
  \citenamefont {Pfeiffer}, \citenamefont {Zervas}, \citenamefont
  {Geiselmann},\ and\ \citenamefont {Kippenberg}}]{Karpov2016a}%
  \BibitemOpen
  \bibfield  {author} {\bibinfo {author} {\bibfnamefont {M.}~\bibnamefont
  {Karpov}}, \bibinfo {author} {\bibfnamefont {H.}~\bibnamefont {Guo}},
  \bibinfo {author} {\bibfnamefont {A.}~\bibnamefont {Kordts}}, \bibinfo
  {author} {\bibfnamefont {V.}~\bibnamefont {Brasch}}, \bibinfo {author}
  {\bibfnamefont {M.~H.~P.}\ \bibnamefont {Pfeiffer}}, \bibinfo {author}
  {\bibfnamefont {M.}~\bibnamefont {Zervas}}, \bibinfo {author} {\bibfnamefont
  {M.}~\bibnamefont {Geiselmann}}, \ and\ \bibinfo {author} {\bibfnamefont
  {T.~J.}\ \bibnamefont {Kippenberg}},\ }\href {\doibase
  10.1103/PhysRevLett.116.103902} {\bibfield  {journal} {\bibinfo  {journal}
  {Physical Review Letters}\ }\textbf {\bibinfo {volume} {116}},\ \bibinfo
  {pages} {103902} (\bibinfo {year} {2016})}\BibitemShut {NoStop}%
\bibitem [{\citenamefont {Porto}\ \emph {et~al.}(1967)\citenamefont {Porto},
  \citenamefont {Fleury},\ and\ \citenamefont {Damen}}]{Porto1967}%
  \BibitemOpen
  \bibfield  {author} {\bibinfo {author} {\bibfnamefont {S.~P.~S.}\
  \bibnamefont {Porto}}, \bibinfo {author} {\bibfnamefont {P.~A.}\ \bibnamefont
  {Fleury}}, \ and\ \bibinfo {author} {\bibfnamefont {T.~C.}\ \bibnamefont
  {Damen}},\ }\href {\doibase 10.1103/PhysRev.154.522} {\bibfield  {journal}
  {\bibinfo  {journal} {Physical Review}\ }\textbf {\bibinfo {volume} {154}},\
  \bibinfo {pages} {522} (\bibinfo {year} {1967})}\BibitemShut {NoStop}%
\bibitem [{\citenamefont {Yang}\ \emph {et~al.}(2016)\citenamefont {Yang},
  \citenamefont {Yi}, \citenamefont {Yang},\ and\ \citenamefont
  {Vahala}}]{Yang2016c}%
  \BibitemOpen
  \bibfield  {author} {\bibinfo {author} {\bibfnamefont {Q.-F.}\ \bibnamefont
  {Yang}}, \bibinfo {author} {\bibfnamefont {X.}~\bibnamefont {Yi}}, \bibinfo
  {author} {\bibfnamefont {K.~Y.}\ \bibnamefont {Yang}}, \ and\ \bibinfo
  {author} {\bibfnamefont {K.}~\bibnamefont {Vahala}},\ }\href {\doibase
  10.1364/OPTICA.3.001132} {\bibfield  {journal} {\bibinfo  {journal} {Optica}\
  }\textbf {\bibinfo {volume} {3}},\ \bibinfo {pages} {1132} (\bibinfo {year}
  {2016})}\BibitemShut {NoStop}%
\bibitem [{\citenamefont {Paschotta}(2004)}]{Paschotta2004}%
  \BibitemOpen
  \bibfield  {author} {\bibinfo {author} {\bibfnamefont {R.}~\bibnamefont
  {Paschotta}},\ }\href {\doibase 10.1007/s00340-004-1548-9} {\bibfield
  {journal} {\bibinfo  {journal} {Applied Physics B}\ }\textbf {\bibinfo
  {volume} {79}},\ \bibinfo {pages} {163} (\bibinfo {year} {2004})}\BibitemShut
  {NoStop}%
\bibitem [{\citenamefont {Haus}\ and\ \citenamefont
  {Mecozzi}(1993)}]{Haus1993}%
  \BibitemOpen
  \bibfield  {author} {\bibinfo {author} {\bibfnamefont {H.~a.}\ \bibnamefont
  {Haus}}\ and\ \bibinfo {author} {\bibfnamefont {A.}~\bibnamefont {Mecozzi}},\
  }\href {\doibase 10.1109/3.206583} {\bibfield  {journal} {\bibinfo  {journal}
  {IEEE Journal of Quantum Electronics}\ }\textbf {\bibinfo {volume} {29}},\
  \bibinfo {pages} {983} (\bibinfo {year} {1993})}\BibitemShut {NoStop}%
\bibitem [{\citenamefont {Matsko}\ \emph {et~al.}(2014)\citenamefont {Matsko},
  \citenamefont {{Wei Liang}}, \citenamefont {Ilchenko}, \citenamefont
  {Savchenkov}, \citenamefont {Byrd}, \citenamefont {Seidel},\ and\
  \citenamefont {Maleki}}]{Matsko2014a}%
  \BibitemOpen
  \bibfield  {author} {\bibinfo {author} {\bibfnamefont {A.~B.}\ \bibnamefont
  {Matsko}}, \bibinfo {author} {\bibnamefont {{Wei Liang}}}, \bibinfo {author}
  {\bibfnamefont {V.~S.}\ \bibnamefont {Ilchenko}}, \bibinfo {author}
  {\bibfnamefont {A.~A.}\ \bibnamefont {Savchenkov}}, \bibinfo {author}
  {\bibfnamefont {J.}~\bibnamefont {Byrd}}, \bibinfo {author} {\bibfnamefont
  {D.}~\bibnamefont {Seidel}}, \ and\ \bibinfo {author} {\bibfnamefont
  {L.}~\bibnamefont {Maleki}},\ }in\ \href {\doibase
  10.1109/IPCon.2014.6995235} {\emph {\bibinfo {booktitle} {2014 IEEE Photonics
  Conference}}}\ (\bibinfo  {publisher} {IEEE},\ \bibinfo {year} {2014})\ pp.\
  \bibinfo {pages} {108--109}\BibitemShut {NoStop}%
\bibitem [{\citenamefont {Matsko}\ \emph {et~al.}(2012)\citenamefont {Matsko},
  \citenamefont {Savchenko},\ and\ \citenamefont {Maleki}}]{Matsko2012}%
  \BibitemOpen
  \bibfield  {author} {\bibinfo {author} {\bibfnamefont {A.}~\bibnamefont
  {Matsko}}, \bibinfo {author} {\bibfnamefont {A.}~\bibnamefont {Savchenko}}, \
  and\ \bibinfo {author} {\bibfnamefont {L.}~\bibnamefont {Maleki}},\
  }\href@noop {} {\bibfield  {journal} {\bibinfo  {journal} {Opt. Lett.}\
  }\textbf {\bibinfo {volume} {37}},\ \bibinfo {pages} {4856} (\bibinfo {year}
  {2012})}\BibitemShut {NoStop}%
\bibitem [{\citenamefont {Yu}\ \emph {et~al.}(2016)\citenamefont {Yu},
  \citenamefont {Jang}, \citenamefont {Okawachi}, \citenamefont {Griffith},
  \citenamefont {Luke}, \citenamefont {Miller}, \citenamefont {Ji},
  \citenamefont {Lipson},\ and\ \citenamefont {Gaeta}}]{Yu2016}%
  \BibitemOpen
  \bibfield  {author} {\bibinfo {author} {\bibfnamefont {M.}~\bibnamefont
  {Yu}}, \bibinfo {author} {\bibfnamefont {J.~K.}\ \bibnamefont {Jang}},
  \bibinfo {author} {\bibfnamefont {Y.}~\bibnamefont {Okawachi}}, \bibinfo
  {author} {\bibfnamefont {A.~G.}\ \bibnamefont {Griffith}}, \bibinfo {author}
  {\bibfnamefont {K.}~\bibnamefont {Luke}}, \bibinfo {author} {\bibfnamefont
  {S.~A.}\ \bibnamefont {Miller}}, \bibinfo {author} {\bibfnamefont
  {X.}~\bibnamefont {Ji}}, \bibinfo {author} {\bibfnamefont {M.}~\bibnamefont
  {Lipson}}, \ and\ \bibinfo {author} {\bibfnamefont {A.~L.}\ \bibnamefont
  {Gaeta}},\ }\href {http://arxiv.org/abs/1609.01760} {\ ,\ \bibinfo {pages}
  {1} (\bibinfo {year} {2016})},\ \Eprint {http://arxiv.org/abs/1609.01760}
  {arXiv:1609.01760} \BibitemShut {NoStop}%
\bibitem [{\citenamefont {Lucas}\ \emph {et~al.}(2016)\citenamefont {Lucas},
  \citenamefont {Karpov}, \citenamefont {Guo}, \citenamefont {Gorodetsky},\
  and\ \citenamefont {Kippenberg}}]{Lucas2016b}%
  \BibitemOpen
  \bibfield  {author} {\bibinfo {author} {\bibfnamefont {E.}~\bibnamefont
  {Lucas}}, \bibinfo {author} {\bibfnamefont {M.}~\bibnamefont {Karpov}},
  \bibinfo {author} {\bibfnamefont {H.}~\bibnamefont {Guo}}, \bibinfo {author}
  {\bibfnamefont {M.}~\bibnamefont {Gorodetsky}}, \ and\ \bibinfo {author}
  {\bibfnamefont {T.}~\bibnamefont {Kippenberg}},\ }\href
  {http://arxiv.org/abs/1611.06567} {\ ,\ \bibinfo {pages} {10} (\bibinfo
  {year} {2016})},\ \Eprint {http://arxiv.org/abs/1611.06567}
  {arXiv:1611.06567} \BibitemShut {NoStop}%
\bibitem [{\citenamefont {Bao}\ \emph {et~al.}(2016)\citenamefont {Bao},
  \citenamefont {Jaramillo-Villegas}, \citenamefont {Xuan}, \citenamefont
  {Leaird}, \citenamefont {Qi},\ and\ \citenamefont {Weiner}}]{Bao2016a}%
  \BibitemOpen
  \bibfield  {author} {\bibinfo {author} {\bibfnamefont {C.}~\bibnamefont
  {Bao}}, \bibinfo {author} {\bibfnamefont {J.~A.}\ \bibnamefont
  {Jaramillo-Villegas}}, \bibinfo {author} {\bibfnamefont {Y.}~\bibnamefont
  {Xuan}}, \bibinfo {author} {\bibfnamefont {D.~E.}\ \bibnamefont {Leaird}},
  \bibinfo {author} {\bibfnamefont {M.}~\bibnamefont {Qi}}, \ and\ \bibinfo
  {author} {\bibfnamefont {A.~M.}\ \bibnamefont {Weiner}},\ }\href
  {http://link.aps.org/doi/10.1103/PhysRevLett.117.163901} {\bibfield
  {journal} {\bibinfo  {journal} {Physical Review Letters}\ }\textbf {\bibinfo
  {volume} {117}},\ \bibinfo {pages} {163901} (\bibinfo {year}
  {2016})}\BibitemShut {NoStop}%
\bibitem [{\citenamefont {Leo}\ \emph {et~al.}(2013)\citenamefont {Leo},
  \citenamefont {Gelens}, \citenamefont {Emplit}, \citenamefont {Haelterman},\
  and\ \citenamefont {Coen}}]{Leo2013}%
  \BibitemOpen
  \bibfield  {author} {\bibinfo {author} {\bibfnamefont {F.}~\bibnamefont
  {Leo}}, \bibinfo {author} {\bibfnamefont {L.}~\bibnamefont {Gelens}},
  \bibinfo {author} {\bibfnamefont {P.}~\bibnamefont {Emplit}}, \bibinfo
  {author} {\bibfnamefont {M.}~\bibnamefont {Haelterman}}, \ and\ \bibinfo
  {author} {\bibfnamefont {S.}~\bibnamefont {Coen}},\ }\href {\doibase
  10.1364/OE.21.009180} {\bibfield  {journal} {\bibinfo  {journal} {Optics
  express}\ }\textbf {\bibinfo {volume} {21}},\ \bibinfo {pages} {9180}
  (\bibinfo {year} {2013})}\BibitemShut {NoStop}%
\bibitem [{\citenamefont {Anderson}\ \emph {et~al.}(2016)\citenamefont
  {Anderson}, \citenamefont {Leo}, \citenamefont {Coen}, \citenamefont
  {Erkintalo},\ and\ \citenamefont {Murdoch}}]{Anderson:16}%
  \BibitemOpen
  \bibfield  {author} {\bibinfo {author} {\bibfnamefont {M.}~\bibnamefont
  {Anderson}}, \bibinfo {author} {\bibfnamefont {F.}~\bibnamefont {Leo}},
  \bibinfo {author} {\bibfnamefont {S.}~\bibnamefont {Coen}}, \bibinfo {author}
  {\bibfnamefont {M.}~\bibnamefont {Erkintalo}}, \ and\ \bibinfo {author}
  {\bibfnamefont {S.~G.}\ \bibnamefont {Murdoch}},\ }\href {\doibase
  10.1364/OPTICA.3.001071} {\bibfield  {journal} {\bibinfo  {journal} {Optica}\
  }\textbf {\bibinfo {volume} {3}},\ \bibinfo {pages} {1071} (\bibinfo {year}
  {2016})}\BibitemShut {NoStop}%
\bibitem [{\citenamefont {Yi}\ \emph {et~al.}(2016{\natexlab{a}})\citenamefont
  {Yi}, \citenamefont {Yang}, \citenamefont {Youl},\ and\ \citenamefont
  {Vahala}}]{Yi2016}%
  \BibitemOpen
  \bibfield  {author} {\bibinfo {author} {\bibfnamefont {X.}~\bibnamefont
  {Yi}}, \bibinfo {author} {\bibfnamefont {Q.-F.}\ \bibnamefont {Yang}},
  \bibinfo {author} {\bibfnamefont {K.}~\bibnamefont {Youl}}, \ and\ \bibinfo
  {author} {\bibfnamefont {K.}~\bibnamefont {Vahala}},\ }\href {\doibase
  10.1364/OL.41.002037} {\bibfield  {journal} {\bibinfo  {journal} {Optics
  Letters}\ }\textbf {\bibinfo {volume} {41}},\ \bibinfo {pages} {2037}
  (\bibinfo {year} {2016}{\natexlab{a}})}\BibitemShut {NoStop}%
\bibitem [{\citenamefont {Jost}\ \emph
  {et~al.}(2015{\natexlab{b}})\citenamefont {Jost}, \citenamefont {Lucas},
  \citenamefont {Herr}, \citenamefont {Lecaplain}, \citenamefont {Brasch},
  \citenamefont {Pfeiffer},\ and\ \citenamefont {Kippenberg}}]{Jost2015b}%
  \BibitemOpen
  \bibfield  {author} {\bibinfo {author} {\bibfnamefont {J.~D.}\ \bibnamefont
  {Jost}}, \bibinfo {author} {\bibfnamefont {E.}~\bibnamefont {Lucas}},
  \bibinfo {author} {\bibfnamefont {T.}~\bibnamefont {Herr}}, \bibinfo {author}
  {\bibfnamefont {C.}~\bibnamefont {Lecaplain}}, \bibinfo {author}
  {\bibfnamefont {V.}~\bibnamefont {Brasch}}, \bibinfo {author} {\bibfnamefont
  {M.~H.~P.}\ \bibnamefont {Pfeiffer}}, \ and\ \bibinfo {author} {\bibfnamefont
  {T.~J.}\ \bibnamefont {Kippenberg}},\ }\href {\doibase 10.1364/OL.40.004723}
  {\bibfield  {journal} {\bibinfo  {journal} {Optics Letters}\ }\textbf
  {\bibinfo {volume} {40}},\ \bibinfo {pages} {4723} (\bibinfo {year}
  {2015}{\natexlab{b}})}\BibitemShut {NoStop}%
\bibitem [{\citenamefont {Papp}\ \emph {et~al.}(2013)\citenamefont {Papp},
  \citenamefont {Del'Haye},\ and\ \citenamefont {Diddams}}]{Papp2013a}%
  \BibitemOpen
  \bibfield  {author} {\bibinfo {author} {\bibfnamefont {S.~B.}\ \bibnamefont
  {Papp}}, \bibinfo {author} {\bibfnamefont {P.}~\bibnamefont {Del'Haye}}, \
  and\ \bibinfo {author} {\bibfnamefont {S.~A.}\ \bibnamefont {Diddams}},\
  }\href {\doibase 10.1103/PhysRevX.3.031003} {\bibfield  {journal} {\bibinfo
  {journal} {Physical Review X}\ }\textbf {\bibinfo {volume} {3}},\ \bibinfo
  {pages} {031003} (\bibinfo {year} {2013})}\BibitemShut {NoStop}%
\bibitem [{\citenamefont {Nozaki}\ and\ \citenamefont
  {Bekki}(1986)}]{Nozaki1986}%
  \BibitemOpen
  \bibfield  {author} {\bibinfo {author} {\bibfnamefont {K.}~\bibnamefont
  {Nozaki}}\ and\ \bibinfo {author} {\bibfnamefont {N.}~\bibnamefont {Bekki}},\
  }\href {\doibase http://dx.doi.org/10.1016/0167-2789(86)90012-6} {\bibfield
  {journal} {\bibinfo  {journal} {Physica D: Nonlinear Phenomena}\ }\textbf
  {\bibinfo {volume} {21}},\ \bibinfo {pages} {381} (\bibinfo {year}
  {1986})}\BibitemShut {NoStop}%
\bibitem [{\citenamefont {Yi}\ \emph {et~al.}(2016{\natexlab{b}})\citenamefont
  {Yi}, \citenamefont {Yang}, \citenamefont {Zhang}, \citenamefont {Yang},\
  and\ \citenamefont {Vahala}}]{Yi2016b}%
  \BibitemOpen
  \bibfield  {author} {\bibinfo {author} {\bibfnamefont {X.}~\bibnamefont
  {Yi}}, \bibinfo {author} {\bibfnamefont {Q.-F.}\ \bibnamefont {Yang}},
  \bibinfo {author} {\bibfnamefont {X.}~\bibnamefont {Zhang}}, \bibinfo
  {author} {\bibfnamefont {K.~Y.}\ \bibnamefont {Yang}}, \ and\ \bibinfo
  {author} {\bibfnamefont {K.}~\bibnamefont {Vahala}},\ }\href
  {http://arxiv.org/abs/1610.08145} {\ ,\ \bibinfo {pages} {1} (\bibinfo {year}
  {2016}{\natexlab{b}})},\ \Eprint {http://arxiv.org/abs/1610.08145}
  {arXiv:1610.08145} \BibitemShut {NoStop}%
\end{thebibliography}%

\end{document}